\documentclass[10pt,twocolumn]{article}
\usepackage[margin=0.75in]{geometry}
\usepackage[T1]{fontenc}
\usepackage{booktabs}
\usepackage{graphicx}
\usepackage{amsmath,amssymb,amsthm}
\newtheorem{proposition}{Proposition}
\usepackage{microtype}
\usepackage[hidelinks]{hyperref}
\usepackage{xcolor}
\newcommand{\drhomacro}{0.911}
\newcommand{\dtransfer}{0.992}
\newcommand{\daddspearman}{0.939}
\newcommand{\daddresid}{17.5\%}
\newcommand{\daddpninety}{48.0\%}
\newcommand{\dsearchverdict}{do not adopt local search into the default method --- the median gain of $0.330$\,pp is below the $0.5$\,pp threshold and 1 cell(s) lost by more than 1\,pp, although 7 of 10 cells did improve}
\newcommand{\dinvall}{9.9\%}
\newcommand{\dinvtoptwentyfive}{15.3\%}
\newcommand{\dinvseed}{12.2\%}
\newcommand{\dbetalo}{0.26}
\newcommand{\dbetahi}{0.77}

\newcommand{\dcalibrsq}{0.84}
\newcommand{\dnmaeiqr}{0.69}

\newcommand{\dheadroomnmae}{0.95}
\newcommand{\dheadroombeta}{-0.76}
\newcommand{\dheadroomn}{10}
\newcommand{\dsearchlow}{1.346}
\newcommand{\dsearchlown}{4}
\newcommand{\dsearchmid}{0.220}
\newcommand{\dsearchmidn}{3}
\newcommand{\dtransfern}{five}
\newcommand{\dcalibdocs}{1{,}500}
\newcommand{\dcalibq}{150}
\newcommand{\dhodocs}{3{,}633}
\newcommand{\dhoq}{323}
\newcommand{\dnconfigslo}{106}
\newcommand{\dnconfigshi}{200}
\newcommand{\dnseededlo}{6}
\newcommand{\dnseededhi}{100}
\newcommand{\dnrandom}{100}
\newcommand{\dnthin}{1}
\newcommand{\dsearchimproved}{7}
\newcommand{\dsearchcells}{10}
\newcommand{\dagreeverdict}{is strongly negative}
\newcommand{\dagreerho}{-0.524}
\newcommand{\dagreetau}{-0.455}
\newcommand{\dagreen}{10}
\newcommand{\dagreebottom}{8}
\newcommand{\dagreeoverlap}{0.13}
\newcommand{\dfpallident}{80.0\%}
\newcommand{\dfpallspread}{11.425}
\newcommand{\dfpalln}{10}
\newcommand{\dfpcollapse}{3}
\newcommand{\dfpn}{7}
\newcommand{\dfpmacro}{0.992}
\newcommand{\dfpident}{78.6\%}
\newcommand{\dfpspread}{0.372}
\newcommand{\dfpqwen}{11.425}
\newcommand{\dfpqwenlo}{4.02}
\newcommand{\dfpqwenhi}{15.44}
\newcommand{\dfpqwenfp}{32.48}
\newcommand{\lieqcanon}{init/L2\_mse/layers}
\newcommand{\dgrantag}{E5-base}
\newcommand{\dgranrho}{0.996}
\newcommand{\dgrancost}{3.7}

\newcommand{\dgrangainsix}{-0.072}
\newcommand{\dgrangaintwelve}{0.397}
\newcommand{\dgrangaintwentyfour}{0.276}

\newcommand{\drandpure}{0.909}
\newcommand{\drandpuremin}{0.872}
\newcommand{\drandpuremax}{0.929}
\newcommand{\drandmixed}{0.892}

\newcommand{\dlabn}{5}
\newcommand{\dlabuniform}{3}
\newcommand{\dlabnonuni}{2}
\newcommand{\dlabrecovered}{3}

\newcommand{\dlabgaplo}{4.257}
\newcommand{\dlabgaphi}{8.765}
\newcommand{\dlabunimodels}{E5-base, EmbGemma, SKILLRET}
\newcommand{\dlabnonunimodels}{BGE-M3 and Qwen3-0.6B}
\newcommand{\ddrn}{5}
\newcommand{\ddrseeds}{3}
\newcommand{\ddrrho}{0.9997}
\newcommand{\ddrham}{0.0}
\newcommand{\ddrident}{100.0\%}
\newcommand{\ddrpass}{5}
\newcommand{\dpacells}{10}
\newcommand{\dparho}{0.9985}
\newcommand{\dpadelta}{0.000}
\newcommand{\dparatio}{10}
\newcommand{\donedimlo}{0.011}
\newcommand{\donedimhi}{0.112}
\newcommand{\dskippedhi}{1.66}
\newcommand{\dctthresh}{0.178}
\newcommand{\dcthigh}{0.407 and 0.325}
\newcommand{\dctnon}{0.032 and 0.013}
\newcommand{\dctcells}{4}
\newcommand{\dctgray}{0}
\newcommand{\dctregretalways}{1.356}
\newcommand{\dctregretnever}{9.928}
\newcommand{\dnullperm}{200}
\newcommand{\dnullrand}{100}
\newcommand{\dnullminperm}{89.5}
\newcommand{\dnullminrand}{100.0}
\newcommand{\dnullpermbeats}{1}
\newcommand{\dnulln}{5}
\newcommand{\dasncells}{4}
\newcommand{\dasnthreshrho}{0.875}
\newcommand{\dasnthreshident}{56.2\%}
\newcommand{\dasnmatchrho}{0.589}
\newcommand{\dasnmatchident}{6.2\%}
\newcommand{\dasnmatchham}{4.0}
\newcommand{\dasnknee}{0.730}
\newcommand{\dlqsignneg}{3}
\newcommand{\dlqsignn}{5}
\newcommand{\dlqsigncollapse}{3}
\newcommand{\dlqsignevery}{3}
\newcommand{\dlqsignarms}{15}
\newcommand{\dlqsignexplains}{coincide exactly}
\newcommand{\dlqcells}{10}
\newcommand{\dlqgraded}{3}
\newcommand{\dlqcliff}{4}
\newcommand{\dlqdead}{3}

\newcommand{\dlqunidead}{5}
\newcommand{\dlqunirescued}{2}
\newcommand{\dlqgapgraded}{0.990}
\newcommand{\dlqgapgradedlo}{0.750}
\newcommand{\dlqgapgradedhi}{7.893}
\newcommand{\dlqgapclifflo}{22.697}
\newcommand{\dlqgapcliffhi}{29.997}
\newcommand{\dlqgemmauni}{0.04}

\newcommand{\dconfhlieq}{3/3}
\newcommand{\dconfhabs}{2/3}
\newcommand{\dconfuni}{0/3}
\newcommand{\dconfunipp}{-0.99, -0.85, -1.01}

\newcommand{\ddevuni}{5/5}
\newcommand{\dpoolhlieq}{8/8}
\newcommand{\dpooluni}{5/8}
\newcommand{\dconfgraded}{2}
\newcommand{\dconfcliff}{2}
\newcommand{\dconfdead}{2}
\newcommand{\dconfrhomin}{0.611}
\newcommand{\dconfrhomax}{0.908}
\newcommand{\dconfrepro}{0.9995}
\newcommand{\dconfabsuni}{1}
\newcommand{\dconfrecovery}{GTE-base at $B_{3}$}
\newcommand{\dconfmxbailow}{23.55}
\newcommand{\dconfmxbaiuni}{5.30}
\newcommand{\dconfmxbaidrift}{28.84}
\newcommand{\dconfmxbaisigned}{27.99}
\newcommand{\dconfgtesigned}{1.21}
\newcommand{\dconfgteabs}{27.51}
\newcommand{\dconfgtesearch}{32.38}
\newcommand{\dconfgteuni}{28.87}

\newcommand{\dconfmpnetsignedret}{0.499}
\newcommand{\dreevaln}{3}
\newcommand{\dreevalmaxd}{0.060}
\newcommand{\dreevalcaseb}{3}
\newcommand{\dreevalcasea}{0}
\newcommand{\dreevalbitsgap}{0.023}
\newcommand{\dreevalcicross}{2}
\newcommand{\dreevalciexcl}{1}
\newcommand{\dreevalnq}{323}
\newcommand{\dremxbaidelta}{-1.04}
\newcommand{\dremxbaicilo}{-1.85}
\newcommand{\dremxbaicihi}{-0.28}

\newcommand{\drempnetdelta}{-0.90}
\newcommand{\drempnetcilo}{-1.90}
\newcommand{\drempnetcihi}{0.18}

\newcommand{\dregtedelta}{-1.01}
\newcommand{\dregtecilo}{-2.71}
\newcommand{\dregtecihi}{0.64}

\newcommand{\dpaqdiff}{two cells}
\newcommand{\dpauntdiff}{two}
\newcommand{\dpauntdelta}{0.79}
\newcommand{\dpoolreq}{176}
\newcommand{\dpoolmin}{154}
\newcommand{\dpoolmax}{174}
\newcommand{\diqorigmax}{0.276}
\newcommand{\diqorigceil}{0.126}
\newcommand{\diqfixmax}{0.358}
\newcommand{\diqfixceil}{0.126}
\newcommand{\diqstdmax}{0.324}
\newcommand{\diqstdceil}{0.157}
\newcommand{\dgtebadtf}{5.14.1}
\newcommand{\dgtegoodtf}{4.49.0}
\newcommand{\dauditn}{146}
\newcommand{\dauditmaxd}{0.0008}
\newcommand{\dauditcross}{none}
\newcommand{\dinfoqxn}{5}

\newcommand{\dinfoqxrungs}{35}
\newcommand{\dinfoqxbestrho}{0.530}
\newcommand{\dinfoqxceil}{6{,}000}
\newcommand{\infoqxoutcome}{It did not satisfy our pre-specified stability criterion up to the original work's 6{,}000-sample scale: changing the projection seed continued to alter a median of 4--9 of the 12 allocation decisions, and no rung produced a byte-identical plan for any seed pair. We therefore do not report unstable NDCG comparisons, in either direction.}
\newcommand{\infoqxnote}{InfoQ-X did not satisfy the pre-specified estimator-stability gate, so no retrieval score is reported for it; see Appendix~\ref{app:infoqx}.}

\date{2026-10-06}
\title{Quantize by Drift: Label-Free Mixed-Precision\\
       Post-Training Quantization for Text Embedders}
\author{Hyojung Han$^{\dagger}$ \qquad Jongmin Kim \qquad Seung-Hun Jeon\\[4pt]
        \small ThakiCloud\\[2pt]
        \small $^{\dagger}$Corresponding author: \texttt{hyojung.han@thakicloud.com}}

\begin{document}
\maketitle

\begin{abstract}
Mixed-precision post-training quantization needs a per-module sensitivity signal, and for a
text embedder the obvious one --- how much retrieval quality a module costs when quantized ---
requires relevance labels that deployments rarely have. We measure a label-free substitute:
\emph{quantization-induced representation drift}, obtained by quantizing one module,
re-encoding the corpus, and recording how far the output embeddings moved from their
full-precision positions. Measuring sensitivity by intervening rather than by inspecting the
full-precision model is not itself new; what is specific here is the quantity observed --- the
displacement of the deployed output representation, which is what a dense retriever ranks
with --- and the setting, encoder-weight mixed precision for retrieval rather than index or
output-vector compression. Across five development embedders, configuration-level drift orders
sampled mixed-precision plans against held-out retrieval quality at a macro Spearman of
\drhomacro, the module sensitivity transports across calibration corpora and retrieval
domains in the usable regime, module drifts compose rank-consistently but not numerically,
and relevance-derived sensitivity adds no consistent value over drift. The method is a single
additive allocation over measured drift under a hard packed-byte budget constraint, with no
labels, no interaction model and no search.
On three embedders held untouched until the method, baselines and hypotheses were frozen
and sealed, the pre-registered directional hypothesis against the prior LieQ criterion holds
(\dconfhlieq\ at the main budget, no collapse), drift scores above the two-sided LieQ steelman in
\dconfhabs, and the label-free signal reproduces under calibration resampling on all three;
but at the main budget the drift allocation is numerically lower than same-budget uniform
precision on all three (\dconfunipp\ points; under independent re-evaluation the paired
query-level interval excludes zero on \dreevalciexcl\ of \dreevaln\ models), and in the one tighter-budget cell where uniform
collapses (\dconfmxbaiuni) both drift (\dconfmxbaidrift) and the LieQ adaptation
(\dconfmxbaisigned) remain usable. We report the confirmation before the pooled numbers and
do not average over it. Together these results position output drift as a robust coarse
sensitivity signal rather than a universally optimal allocation objective: it avoids the
catastrophic failures of the transferred signed-geometry adaptation and can remain usable at
stressed budgets where uniform collapses, but fine-grained redistribution around a strong
uniform operating point remains unresolved.
\end{abstract}

\section{Introduction}

Quantizing a text embedder is not the same problem as quantizing a generative model. Dense
retrievers encode queries and documents independently and rank by a similarity between the two
vectors~\cite{karpukhin2020dpr,xiong2021ance,izacard2022contriever}, and a single embedder is
typically served behind every query a system answers, which makes it a natural target for
low-bit weight compression. The
output is a vector whose \emph{geometry} is the product, and the downstream metric --- ranking
quality --- depends on relative distances rather than on any single token. Mixed-precision
allocation therefore needs a sensitivity signal that tracks retrieval damage, and the
tempting choice is to measure that damage directly with relevance judgements. Deployments
usually do not have them.

This paper reports what happens when the signal is label-free and produced by the
quantization itself. For each module and bit width we apply the quantization, re-encode the
calibration corpus, and measure the cosine displacement of the resulting embeddings from
their full-precision positions.

\paragraph{What the claim is, and what it is not.} A reader can reasonably object that
``embeddings that move less are embeddings that rank better'' is close to a tautology, and
that finding a correlation between whole-model drift and retrieval quality is not a
contribution. We agree, and that is not the claim. The unit of novelty here is not the
correlation but the \emph{allocation signal}: whether drift measured by quantizing a
\emph{single module in isolation} carries enough information to decide that module's
precision, and whether the resulting ordering is stable enough across models, budgets and
retrieval domains to be used without any downstream supervision. Those are empirical
questions with a real chance of a negative answer --- single-module drift could compose too
badly to predict a configuration, it could be dominated by module size, or it could reorder
whenever the calibration corpus changes. \S\ref{sec:composition} and \S\ref{sec:transfer}
test each of them.

\paragraph{Two jobs for a sensitivity signal.} The results separate two things a
sensitivity signal is asked to do. One is to avoid catastrophic precision mistakes --- to
keep bits away from a module whose quantization collapses retrieval and to survive budgets
at which the trivial allocation does not. The other is to resolve small trade-offs among
allocations that are already good. Drift is much stronger at the first than at the second,
and the paper's own development analysis already says why: whole-configuration drift orders
sampled allocations against held-out quality at a macro Spearman of \drhomacro\ over the
full pool, but the correlation falls markedly once the pool is restricted to its best quarter
(\S\ref{sec:predict}, Table~\ref{tab:stratified}), and the inversion rate of the additive prediction
rises in the same region (\S\ref{sec:composition}). The confirmation on three untouched
models (\S\ref{sec:confirm}) is the same fact seen from the other side: the pre-registered
comparison against a prior geometry criterion replicates, and at the budget where
uniform precision is already healthy, drift does not improve on it. We treat that not as a
failed method but as the boundary of what a coarse surrogate can be expected to resolve. The
position this paper ends in is therefore narrower than the one it started from: output drift
is a robust coarse signal for avoiding harmful precision assignments across heterogeneous
embedders, and it is not a universally superior fine-grained allocator --- on untouched
models it scores above a prior geometry-based adaptation while consistently, if not
always significantly, trailing uniform precision at the healthy 3-bit operating point.

\paragraph{Relation to intervention-based allocation.} Measuring sensitivity by actually
quantizing a component is an established idea. InfoQ~\cite{akbulut2026infoq} quantizes each
layer at each candidate bit width and scores it by the resulting change in mutual information
in the \emph{subsequent layers}, then allocates by integer linear programming; our allocator
is the same integer program over summed per-module scores, and we make no claim of novelty for
it. The difference is in what is observed. InfoQ reads an internal, information-theoretic
quantity --- sliced mutual information between activations at downstream observer points ---
inside a vision CNN, and validates on ImageNet accuracy. We read the \emph{final output
embedding}, which for a dense retriever is not one representation among many but the exact
vector the ranking score is computed from; Proposition~\ref{prop:bound} makes that connection
quantitative. Whether that difference in observable changes the allocation, and changes it
for the better on retrieval, is an empirical claim that \S\ref{sec:confirm} settles only
in part: against the prior geometry criterion, yes; against uniform precision at a
healthy operating point, no. Two further differences are matters of evaluation rather than method:
we measure how far the sensitivity transports across retrieval domains, and we express
budgets in real packed bytes including scale and zero-point metadata.
\S\ref{sec:related} places both against geometry-based and interaction-aware allocators.

\paragraph{Contributions.}
\begin{itemize}
  \item \textbf{A retrieval-facing, label-free sensitivity signal} (\S\ref{sec:method}).
        Quantization-induced displacement of the output embedding, measured per module and
        bit width, requiring no relevance judgements, no gradients and no search.
        Proposition~\ref{prop:bound} connects it deterministically to the perturbation of the
        cosine score a dense retriever ranks with.
  \item \textbf{An empirical study of output-drift allocation across heterogeneous text
        embedders and actual packed-byte regimes} (\S\ref{sec:setup}--\S\ref{sec:results}),
        including how far module sensitivity transports across retrieval domains, and a
        pre-registered confirmation on three untouched embedders (\S\ref{sec:confirm}) that
        characterises the signal's strengths and its limits rather than only its wins: the
        comparison against a prior geometry criterion replicates, the comparison against
        uniform precision at the main budget does not. The allocator itself is the standard
        additive integer program and is not claimed as a contribution.
  \item \textbf{A mechanistic account of why the allocator can be this simple}
        (\S\ref{sec:composition}--\S\ref{sec:analysis}). Module drift composes
        rank-consistently but not numerically; relevance supervision and local
        configuration search each fail to add consistent value, and we measure why.
\end{itemize}

\section{Method}\label{sec:method}

\paragraph{Drift.} Let $E_0$ be the full-precision embeddings of the calibration corpus and
$E_{i,b}$ the embeddings after quantizing module $i$ to $b$ bits. The sensitivity of that cell
is the mean cosine displacement
\begin{equation}\label{eq:drift}
\begin{aligned}
  d_{i,b} \;=\; \frac{1}{2}\Big[\,
    &\frac{1}{|Q|}\sum_{q \in Q}\big(1 - \langle E_{i,b}(q),\, E_0(q)\rangle\big) \\
    &+ \frac{1}{|D|}\sum_{d \in D}\big(1 - \langle E_{i,b}(d),\, E_0(d)\rangle\big)\Big] ,
\end{aligned}
\end{equation}
over the calibration documents $D$ and queries $Q$, with every embedding re-encoded through
the quantized model. No relevance judgement enters this quantity.

The two sides are weighted equally rather than by their counts. An earlier form of this
definition pooled them, which on our document-heavy corpora ($|D|{:}|Q| = \dparatio{:}1$)
made the sensitivity predominantly a document-drift statistic; the margin bound below charges
the query and document sides equally, so the balanced form is the one the theory points to.
We fixed this on the development models before opening the confirmation models, and audited
what the change costs: measuring the module drift once with the two sides kept separate and
building all four poolings from that single measurement, the balanced, pooled and
document-only variants \textbf{selected the same bit plan in all \dpacells\ development
cells} (the query-only variant differed in \dpaqdiff), the balanced and pooled allocations
differed by \dpadelta\ NDCG points, and the sensitivity rankings correlated at
$\ge \dparho$ (Appendix~\ref{app:pooling}). We therefore adopted the theoretically motivated
form without discarding any measurement. On the untouched models the two forms were no
longer interchangeable at the tighter budget: on \dpauntdiff\ of the three the balanced plan
differs from the pooled one at $B_{2\leftrightarrow3}$, by up to \dpauntdelta\ NDCG points in
the balanced form's favour, which is reported in Appendix~\ref{app:pooling}; the frozen
method is the balanced form throughout.

\paragraph{Why this observable, and not an internal one.} Cosine drift on the output is not
merely a convenient proxy: for a retriever that scores by inner product on $\ell_2$-normalized
embeddings, it bounds the perturbation of the score itself. Write $\delta(x) = 1 -
\langle E_Q(x), E_0(x)\rangle$ for the per-item drift in Equation~\ref{eq:drift}, and
$s(q,d) = \langle E(q), E(d)\rangle$ for the retrieval score.

\begin{proposition}\label{prop:bound}
For unit-norm embeddings, $\|E_Q(x) - E_0(x)\|_2 = \sqrt{2\delta(x)}$, and hence
\begin{equation}\label{eq:score-bound}
  |s_{0}(q,d) - s_{Q}(q,d)| \;\le\; \sqrt{2\delta(q)} + \sqrt{2\delta(d)} .
\end{equation}
For the ranking margin $m(q,p,n) = s(q,p) - s(q,n)$ between a positive $p$ and a negative $n$,
\begin{equation}\label{eq:margin-bound}
  |m_{0} - m_{Q}| \;\le\; 2\sqrt{2\delta(q)} + \sqrt{2\delta(p)} + \sqrt{2\delta(n)} .
\end{equation}
\end{proposition}

\noindent The proof is two lines of Cauchy--Schwarz and is given in
Appendix~\ref{app:proof}. We state it as motivation, not as a guarantee. It bounds the
distortion of the \emph{score}, not the number of ranking inversions; a margin can be small
enough that an arbitrarily small perturbation flips it, and our signal is a corpus
\emph{mean} whereas Equations~\ref{eq:score-bound}--\ref{eq:margin-bound} are per-item, so
Jensen's inequality only gives $\mathbb{E}[\sqrt{2\delta}] \le \sqrt{2\bar{d}}$ in one
direction. What it does establish is that the quantity we measure is the one the retrieval
score is built from, with a deterministic connection between the two --- which is the sense in
which this observable is retrieval-specific rather than a generic representation-quality
statistic.

Equation~\ref{eq:margin-bound} also has a design consequence we did not originally act on.
Per triple it charges weight $2$ to the query and weight $1$ to each document, so the query
and document sides together carry equal weight. That is the argument for the balanced form of
Equation~\ref{eq:drift}, and the decision was pre-registered on the three axes above ---
ranking, allocation and held-out score --- and resolved on the development models before the
confirmation models were opened, not after seeing which variant scored better.

\paragraph{Allocation.} The plan minimises $\sum_i d_{i,b_i}$ subject to a hard packed-byte
budget constraint $C(c) \le B$, where $C(c)$ is the realised packed cost of plan $c$ including
group metadata; the solver never exceeds $B$ and may land slightly below it. This is the same integer program used by intervention-based allocators for other
domains~\cite{akbulut2026infoq}; the contribution is the signal fed into it, not the solver.
It is deliberately the simplest thing that could work: no interaction term, no search, no
labels. \S\ref{sec:analysis} reports the measurements that removed each of the alternatives we
started with.

\section{Experimental setup}\label{sec:setup}

\paragraph{Models.} Five development embedders: E5-base-v2~\cite{wang2022e5}, BGE-M3~\cite{chen2024bgem3},
EmbeddingGemma-300M~\cite{vera2025embeddinggemma},
Qwen3-Embedding-0.6B~\cite{zhang2025qwen3embedding} and SKILLRET-0.6B, abbreviated in the tables as E5-base, BGE-M3,
EmbGemma, Qwen3-0.6B and SKILLRET. Three further embedders --- mxbai-embed-large-v1,
all-mpnet-base-v2~\cite{reimers2019sbert,song2020mpnet} and
gte-base-en-v1.5~\cite{li2023gte,zhang2024mgte}, abbreviated mxbai-large, MPNet-base and GTE-base ---
were held out entirely until the method, the baselines, the budget grid, every threshold and
the hypotheses were frozen and their pre-registration sealed by hash; they were then run once
under the frozen protocol and are reported in \S\ref{sec:confirm} under the pre-registered
order, before any pooled number.

\paragraph{Quantizer.} Weight-only group-wise asymmetric round-to-nearest~\cite{gholami2021survey} with a per-group
FP16 scale and zero point; activations are left in full precision. Candidate bit widths are
$\{2,3,4\}$ per module and tensors with fewer than two dimensions are not quantized. Those
one-dimensional parameters (biases and normalisation weights) are excluded from both the
numerator and the denominator of the packed-bit accounting; across the eight models they are
\donedimlo--\donedimhi\% of all parameters, and two-dimensional tensors that are kept in full
precision are charged at 16 bits and amount to at most \dskippedhi\% of parameters. Both rules
apply identically to every arm of a model. The
group size is chosen per model so that a uniform 3-bit packing is a feasible point of the
budget grid, and therefore differs across models; Table~\ref{tab:setup} reports it together
with the realized packed cost of each budget.

\paragraph{Retrieval.} Documents and queries are encoded with each model's own pooling and
prompt conventions through \texttt{sentence-transformers}~\cite{reimers2019sbert}, the resulting embeddings are
$\ell_2$-normalized, and retrieval scores are inner products on those unit vectors --- that
is, cosine similarity. Search is exact rather than approximate, so no index-side
approximation enters the measurement. This is the setting Proposition~\ref{prop:bound}
assumes. ``Full precision'' throughout means the released weights loaded in FP32; no
half-precision inference is used anywhere in the pipeline, and every quantized arm is
produced by round-to-nearest from that same FP32 copy, so the label applies identically to
every arm. (Earlier drafts wrote FP16 for this reference; the accounting charges skipped
tensors at 16 bits, which is a deployment assumption, not the evaluation dtype.) Differences
quoted in the text are computed from unrounded ledger values, so subtracting two displayed
two-decimal cells can differ from the quoted difference by $0.01$. One software asymmetry is disclosed here rather than hidden in a footnote:
the remote modelling code of \texttt{gte-base-en-v1.5} indexes its rotary position table with
uninitialised positions under \texttt{transformers}~\dgtebadtf\ and fails before producing an
embedding, so that model alone is evaluated under \texttt{transformers}~\dgtegoodtf, with every arm
for that model --- full precision, uniform, both LieQ variants and drift --- under the same pinned
version; the other seven models run under release \dgtebadtf, which the container resolves by default.
The diagnostic that isolated the version is in the released ledgers.

\begin{table*}[t]\centering\footnotesize
\begin{tabular}{lccccc}
\toprule
 & & & \multicolumn{2}{c}{Realized packed bpw} & \\
\cmidrule(lr){4-5}
Model & Group & Pool & $B_{2\leftrightarrow3}$ & $B_{3}$ & Pred.\ pool budget \\
\midrule
E5-base & 64 & 174 & 3.117 & 3.612 & 3.612 ($=S_3$) \\
BGE-M3 & 32 & 154 & 3.708 & 4.200 & 4.200 ($=S_3$) \\
EmbGemma & 128 & 174 & 2.750 & 3.250 & 3.750 \\
Qwen3-0.6B & 32 & 174 & 3.500 & 4.000 & 4.000 ($=S_3$) \\
SKILLRET & 32 & 174 & 3.500 & 4.000 & 4.000 ($=S_3$) \\
\bottomrule
\end{tabular}

\caption{Per-model configuration. ``Group'' is the quantization group size, ``Pool'' the
realized number of distinct feasible configurations used for the predictability analysis
after deduplication, and the two budget columns give the \emph{actual} packed bits per
weight that each budget name resolves to on that model. The budget names are not bit widths:
$B_{2\leftrightarrow3}$ resolves to a different packed cost on every model.}
\label{tab:setup}
\end{table*}

\paragraph{Allocation units.} We partition the encoder into twelve units: four attention
layer-bands, four feed-forward layer-bands, and four vocabulary bands of the embedding table.
Layer bands are equal splits of the layer index, so band $j$ holds layers
$\lfloor jL/4 \rfloor \ldots \lfloor (j{+}1)L/4 \rfloor - 1$ for an $L$-layer encoder; this
makes the partition comparable across models with different depths. Vocabulary bands are
\emph{contiguous row slices of the embedding matrix in tokenizer-ID order}, that is, rows
$\lfloor jV/4 \rfloor \ldots \lfloor (j{+}1)V/4 \rfloor - 1$ of a $V$-row table. The partition
is therefore a fixed property of the released tokenizer and is \emph{not} derived from token
frequencies in our calibration corpus, so the partition itself does not introduce an
additional corpus-dependent grouping rule. It does not follow that the embedding bands are
domain-invariant: different corpora exercise different token IDs, so the measured sensitivity
of a band can still depend on the domain. That is an empirical question, and
\S\ref{sec:transfer} is where it is answered rather than assumed. Bands rather than individual
layers keep the configuration
space enumerable while still letting the allocator separate early from late computation and
low-ID from high-ID vocabulary. Deploying per-band precision on the embedding table requires a
row-blocked packing format; we account for its scale and zero-point overhead in the budget
below but do not report kernel-level latency. Whether twelve is the right number of units is
checked in Appendix~\ref{app:granularity}.

\paragraph{Budget.} Budgets are expressed in \emph{actual packed bits per weight}, including
the group-wise scale and zero-point overhead, because a nominal ``3.5\,bpw'' means a different
memory footprint on models with different group sizes and padding. Writing $S_b$ for the
packed cost of a uniform $b$-bit model, we evaluate at two operating points,
$B_{2\leftrightarrow3}=(S_2{+}S_3)/2$ and $B_{3}=S_3$: a stressed regime in which uniform
precision is often unusable, and the healthy regime in which it is. A third point,
$B_{3\leftrightarrow4}=(S_3{+}S_4)/2$, appeared in early drafts of the protocol; it was never
measured on any model, the frozen confirmation pre-registration names only the two points
above, and the scope of the paper is fixed to them. We keep the older labels $B_{2.5}$ and
$B_{3.0}$ in the tables for continuity with the ledger, but they are budget \emph{names}, not bit widths: the realized packed cost of
$B_{2.5}$ differs across models (Table~\ref{tab:frontier} reports it per cell), and reading
``2.5'' as 2.5\,bpw is wrong.

\paragraph{Corpora.} All corpora are taken from BEIR~\cite{thakur2021beir}, whose
heterogeneous zero-shot tasks also make up most of the retrieval portion of MTEB~\cite{muennighoff2023mteb}.
Module drift is profiled on a probe corpus (BEIR ArguAna) subsampled to
\dcalibdocs\ documents and \dcalibq\ queries; all reported retrieval quality is on a disjoint
held-out corpus (BEIR NFCorpus, \dhodocs\ documents and \dhoq\ queries). Probe and evaluation banks are
disjoint throughout, and no relevance judgement from either corpus enters the drift
computation.

\paragraph{Configuration pools.} How the pool is built matters, because a correlation measured
on it inherits its composition. We request 176 plans per model and keep the distinct feasible
ones, which leaves the per-model counts in Table~\ref{tab:setup}. The requested mix is 50\%
one- and two-swap perturbations of the uniform plan, 25\% perturbations of allocator outputs
(the drift plan and its neighbourhood), and 25\% plans drawn uniformly at random from the
feasible set at the same budget. The allocator-derived quarter means the pool is \emph{not}
independent of the signal being evaluated. We regard the pooled $\rho$ as
descriptive of this pool rather than of the feasible set, and \S\ref{sec:limits} records a
drift-independent random-only pool as required before the number is used as a headline. For
the search and composition analyses each cell evaluates \dnrandom\ random feasible plans plus
every distinct feasible neighbour of the seeded plan, and the two groups are reported
separately in Table~\ref{tab:composition}. The seeded neighbourhood is not the same size in
every cell: it ranges from \dnseededlo\ to \dnseededhi\ plans, so a cell holds between
\dnconfigslo\ and \dnconfigshi\ configurations in total. \dnthin\ cell has fewer than fifty
seeded neighbours, and its $\Delta_{\mathrm{search}}$ is correspondingly a statement about a
small neighbourhood rather than a broad one; Table~\ref{tab:headroom} gives the per-cell
counts.

\section{Results}\label{sec:results}

\subsection{Drift predicts retrieval damage}\label{sec:predict}

\begin{table*}[t]\centering\footnotesize
\begin{tabular}{lcccc}
\toprule
Model & Budget on grid & $\rho$ & $\tau_b$ & Top-10\% recall \\
\midrule
E5-base & yes & 0.892 & 0.720 & 0.41 \\
BGE-M3 & yes & 0.973 & 0.865 & 0.73 \\
EmbGemma & no & 0.891 & 0.738 & 0.65 \\
Qwen3-0.6B & yes & 0.916 & 0.769 & 0.53 \\
SKILLRET & yes & 0.884 & 0.732 & 0.59 \\
\midrule
Macro ($n=5$) & --- & 0.911 & 0.765 & 0.58 \\
\bottomrule
\end{tabular}

\caption{Spearman $\rho$ and Kendall $\tau_b$ between whole-model embedding drift and held-out
NDCG@10 over a pool of up to \dpoolreq\ requested mixed-precision configurations per model
(\dpoolmin--\dpoolmax\ distinct feasible plans after deduplication), and the recall
of the true top decile by the drift ranking. ``Budget on grid'' marks cells whose frozen budget
coincides with that model's uniform 3-bit packed size, which constrains the feasible set.}
\label{tab:predictability}
\end{table*}

\begin{figure}[t]\centering
\includegraphics[width=\columnwidth]{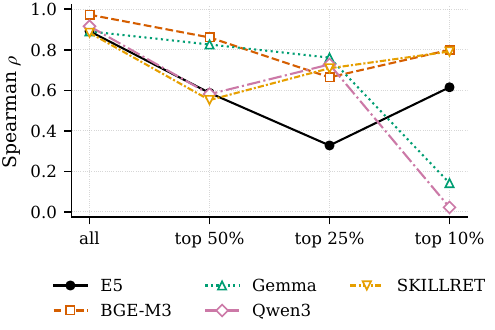}
\caption{Spearman correlation between whole-model drift and held-out NDCG@10, restricted to
progressively better configurations. ``top 25\%'' means the quarter of the pool with the
highest true NDCG. The correlation over the full pool is carried in part by collapsed
configurations; among good configurations it is substantially weaker. The top-decile column
has $n\approx16$ per model and is shown for shape only.}
\label{fig:predictability}
\end{figure}

Configuration-level drift orders mixed-precision plans well
(Table~\ref{tab:predictability}), but the headline number needs the qualification in
Figure~\ref{fig:predictability}: restricted to the better half of the pool the correlation
drops, and restricted to the better quarter it drops again (Table~\ref{tab:stratified}).
Drift is an excellent coarse surrogate for retrieval quality and a much weaker fine one.
\S\ref{sec:analysis} shows that this distinction has a direct algorithmic consequence.

\subsection{Allocation quality against budget-matched baselines}\label{sec:against}

\begin{table*}[t]\centering\footnotesize
\begin{tabular}{lcccccc}
\toprule
Model & FP32 & Uniform & LieQ-signed & LieQ-abs & InfoQ-X & Drift (ours) \\
\midrule
E5-base & 34.58 & 31.24 & 24.40 & 31.24$^{\equiv\mathrm{U}}$ & rejected$^{\ddagger}$ & 32.29 \\
BGE-M3 & 31.50 & 27.67 & 0.96$^{\S}$ & 21.38 & rejected$^{\ddagger}$ & 30.15 \\
EmbGemma & 38.52 & 36.13 & 36.13 & 36.13$^{\equiv\mathrm{U}}$ & rejected$^{\ddagger}$ & 37.12 \\
Qwen3-0.6B & 34.97 & 31.00 & 1.31$^{\S}$ & 27.05 & rejected$^{\ddagger}$ & 31.31 \\
SKILLRET & 33.74 & 29.01 & 1.20$^{\S}$ & 29.01$^{\equiv\mathrm{U}}$ & rejected$^{\ddagger}$ & 30.62 \\
\bottomrule
\end{tabular}

\caption{\textbf{Development comparison} (the five models the method was designed against;
the three untouched confirmation models are reported separately in \S\ref{sec:confirm},
and no aggregate win count against LieQ is given over these five). Held-out NDCG@10 at $B_{3}$, all
arms under the same packed-byte budget. $^{\S}$marks a cliff-crossing cell, where one allocation falls
below the usability threshold, so the difference is not a quality margin.
\textbf{LieQ-signed} is the literal port of the prior directional criterion;
\textbf{LieQ-abs} replaces its score with $|\Delta r|$ and is a two-sided steelman fixed in
advance of computing it, after a systematic sign reversal was observed on the development
models. $^{\equiv\mathrm{U}}$marks a LieQ-abs allocation that is byte-identical to uniform
precision --- in those cells the unsigned criterion does not discriminate between modules, and
the drift margin should be read as a margin over uniform rather than over LieQ.
$^{\ddagger}$InfoQ-X was rejected by a pre-specified estimator-stability gate before any
retrieval score was computed --- an empty cell here is a resolved negative result, not a
missing measurement. LieQ-adapt uses
LieQ's own diagnostic~\cite{lieq2025} --- the training-induced concentration of a
projection's output spectrum, measured against an untrained weight matrix of matched
statistics --- as the sensitivity score, with the quantizer, bit set, group size, packing
overhead, budget, allocator and calibration inputs held identical to ours. It is an adaptation
rather than a reproduction: LieQ allocates per layer, we allocate over twelve units. Fifteen arms
are run, and the one reported here is fixed \emph{a priori} as closest to the method as described,
not chosen on the held-out score; the arm and the selection rule are given in
Table~\ref{tab:lieqarms}, and the post-hoc best of the fifteen is reported separately as a
LieQ-family upper bound in Table~\ref{tab:lieqoracle}. InfoQ-X is reported only if it satisfies the
estimator-stability criterion fixed in advance (Appendix~\ref{app:infoqx});
\infoqxnote{} Our own static-geometry operationalization is not a published system and is
reported in Appendix~\ref{app:tables} (Table~\ref{tab:staticgeo}) rather than here.}
\label{tab:baselines}
\end{table*}

\begin{table}[t]\centering\footnotesize
\begin{tabular}{lcccc}
\toprule
Model & Arm & Gain (pp) & Seed $\rho$ & Agreement $\rho$ \\
\midrule
E5-base & init/$4^{-b}$/lay & -6.842 & 1.000 & -0.762 \\
BGE-M3 & init/$4^{-b}$/lay & -26.708 & 1.000 & -0.497 \\
EmbGemma & init/$4^{-b}$/lay & 0.000 & 0.976 & -0.427 \\
Qwen3-0.6B & init/$4^{-b}$/lay & -29.697 & 0.976 & -0.524 \\
SKILLRET & init/$4^{-b}$/lay & -27.813 & 1.000 & -0.545 \\
\bottomrule
\end{tabular}

\caption{Which LieQ-adapt arm is reported and how reproducible its score is.
\textbf{The reported arm is fixed in advance, not selected on the held-out score.} Fifteen
arms are run --- three sensitivity variants crossed with five allocation rules, the latter
being LieQ's own top-$m$ four-/two-bit split at a matched packed budget plus two bit-penalty
forms crossed with two scopes --- and the arm in Table~\ref{tab:baselines} is the one chosen
\emph{a priori} as closest to the method as described: an isotropic untrained reference, LieQ's own per-head split, LieQ's criterion applied
to the layer bands with the vocabulary bands held at uniform (LieQ is layer-wise and has no
vocabulary-band notion), and the $4^{-b}$ bit penalty that matches uniform-quantization error
scaling. Choosing instead the best of the fifteen by held-out NDCG would be a
\emph{LieQ-family post-hoc upper bound}, not a baseline; that quantity is reported separately
in Appendix~\ref{app:tables} (Table~\ref{tab:lieqoracle}) together with the arm that uses
LieQ's own two-level allocation rule, and our method is given no comparable advantage. Arm
names are abbreviated: the prefix is the sensitivity variant (\texttt{init} = isotropic
untrained reference, \texttt{w}$d$ = every projection split at a fixed width $d$,
\texttt{colmatch} = a reference matched to the trained weight's column statistics),
\texttt{top-$m$} is LieQ's own allocation rule, and $2^{-b}$/$4^{-b}$ with
\texttt{all}/\texttt{lay} name the bit-penalty form and whether embedding bands participate.
These are display abbreviations for one ledger key and not separate recipes: the reported arm
is written \texttt{\lieqcanon} in the ledger and renders here as
\texttt{init}/$4^{-b}$/\texttt{lay}, where \texttt{L2\_mse} is the $4^{-b}$ penalty --- the
error of a uniform quantizer scales as $4^{-b}$, so the two names denote the same
quantity.
Seed $\rho$ is the rank correlation of the module scores across draws of the untrained
reference matrix, which is the only stochastic element of the diagnostic. Agreement $\rho$
is the rank correlation between the two sensitivity criteria over the twelve allocation
units; since the allocator, bit set, group size, packing overhead, budget and calibration
inputs are identical between the arms, it isolates the disagreement between criteria from
everything else.}
\label{tab:lieqarms}
\end{table}

\paragraph{Why the InfoQ-X column is empty.} The second published criterion we tried is
absent from Table~\ref{tab:baselines} for a reason worth stating in the body rather than only
in an appendix. An allocator consumes the \emph{ordering} of module sensitivities, so we fixed
an estimator-reproducibility criterion in advance and resolved it before computing any
retrieval score. The InfoQ-X adaptation did not meet it: over \dinfoqxrungs\ model-rungs --- seven
ladder settings on each of \dinfoqxn\ models --- reaching the original work's
\dinfoqxceil-sample scale, the seed-pairwise rank correlation of the sensitivity scores never
rose above \dinfoqxbestrho, against the $0.9$ fixed in advance. \textbf{The adaptation was
therefore rejected before outcome evaluation, not left unrun}, and we do not report an NDCG
comparison that would be a comparison of noise. The allocations also never reproduced, but we
do \emph{not} treat that as the load-bearing evidence: Appendix~\ref{app:allocsens} shows that
a known-good signal degraded to the same rank agreement also almost never reproduces an
allocation, so plan identity at this level of agreement says more about the allocator than
about the estimator. Appendix~\ref{app:infoqx} gives the criterion and the ladder, and
Appendix~\ref{app:audit} the implementation audit that qualifies it.

\paragraph{And our own signal under an analogous audit.} A gate that rejects a baseline
should also be one the proposed method is asked to face. Drift has no stochastic element
inside the estimator --- given a calibration corpus it is deterministic --- so there is no
exact analogue of redrawing a projection bank. We therefore varied the one stochastic input it
does have, resampling which \ddrseeds\ calibration subsets the sensitivity is measured on, and
computed the same three statistics the stability ladder reports. On all \ddrn\ development
models the seed-pairwise rank correlation is \ddrrho, the median plan Hamming distance is
\ddrham, and the selected plan is byte-identical in \ddrident\ of seed pairs; \ddrpass\ of
\ddrn\ models satisfy the criterion that InfoQ-X failed (Table~\ref{tab:driftrepro}).
\textbf{The two randomness sources are different}, so these numbers are not a controlled
comparison of estimator variance --- resampling a calibration subset perturbs less than
redrawing a projection bank, and the audit is correspondingly easier. What it does establish
is that the allocation our method ships is not an accident of which calibration examples were
drawn.

\paragraph{Graded margins and cliff crossings are different results.} A quantization
budget can put a model on either side of a utility cliff, so a gap in NDCG can mean two
unrelated things. We therefore classify every cell by the usability of each arm on its own,
using the same threshold as the domain-transport analysis: an allocation is usable when it
retains at least half of full-precision quality. We call the result the cell's
\emph{pairwise comparison stratum}, and it is a different quantity from the
\emph{drift operating regime} in Table~\ref{tab:frontier}, which classifies a cell by the
drift allocation alone and says nothing about the baseline. A cell can sit in a useful drift
operating regime and still be a cliff crossing, which is exactly what BGE-M3 does at $B_{3}$.
In a \emph{graded} cell both allocations are
usable and the gap is a quality margin. In a \emph{cliff-crossing} cell exactly one allocation
is usable and the gap measures which side of the cliff the allocation landed on. In a
\emph{dead} cell neither is usable and the budget, not the criterion, is what failed. We report
all three and never average a cliff crossing into a graded margin.

Of the \dlqcells\ measured cells, \dlqgraded\ are graded, \dlqcliff\ are cliff crossings and
\dlqdead\ are dead. In the graded stratum the margin is \dlqgapgradedlo\ to \dlqgapgradedhi\
points, median \dlqgapgraded. \textbf{Three graded development cells are insufficient for an
aggregate comparative claim}; they are stated in full rather than averaged into one favourable
number, and they are why the three untouched confirmation models are held back. The cliff-crossing stratum is a different and, we think,
more interesting finding. In every one of them the drift allocation is the usable one and the
LieQ allocation is not; the reverse never occurs. Read as margins those cells say
\dlqgapclifflo\ to \dlqgapcliffhi\ points, which is why we do not read them as margins. Read
correctly they say that \emph{the choice of sensitivity criterion can decide whether a
same-budget allocation stays on the usable side of a quantization cliff.}

We can say what the collapses run through, and it is not arm selection. On \dlqsignevery\ of
\dlqsignn\ development models \emph{every one} of the \dlqsignarms\ arms falls below the
threshold, including the post-hoc best and LieQ's own two-level allocation rule. All
\dlqsignarms\ share one directional reading of the score, so this shows that no choice of
\emph{allocation rule} repairs the collapse --- not that every reasonable reading of LieQ's
criterion fails. The immediate failure mode is the sign of the
transferred score. LieQ's $\Delta r$ measures how much more concentrated a trained
projection's output spectrum is than an untrained one, and its allocation rule treats a
larger value as more sensitive. In these retrieval encoders the attention output projections
give \emph{negative} $\Delta r$ --- the trained projection's spectrum is flatter than the
random reference --- so read literally the criterion ranks every attention band least
sensitive and the allocator gives them all the minimum width. The set of models whose attention
bands are mostly negative and the set whose LieQ allocation collapses \dlqsignexplains\ (\dlqsignneg\ and \dlqsigncollapse\ models, respectively). That is an association across five
models, not yet a demonstrated cause; the two-sided steelman below tests it
directly by re-running the criterion with the direction removed.

The association alone could not distinguish two explanations --- a criterion that genuinely
fails to transfer, or a directional assumption meeting a sign the original setting never
produced --- so we tested it by intervention. We fixed a single additional score,
$S = |\Delta r|$, before computing it: the untrained reference, the calibration inputs, the
unit partition, the budget, the bit penalty and the allocator are all held at the \emph{signed}
arm's values, and the definition together with a ban list on further variants was sealed as an
amendment to the confirmation pre-registration. We call it \textbf{LieQ-abs}, and the literal
port \textbf{LieQ-signed}.

Removing the direction removes the catastrophic failures. \dlabrecovered\ of the collapsed
cells return to usable quality, so those failures were \emph{caused} by transferring LieQ's
directional reading of spectral concentration to encoders exhibiting negative $\Delta r$.
\textbf{This does not show that LieQ's underlying geometry is invalid for retrieval}; the
unsigned steelman substantially recovers the failed models.

What it recovers to splits in two, and the split matters more than any average over it. On
\dlabuniform\ of \dlabn\ models --- \dlabunimodels\ --- the recovered allocation is
\emph{byte-identical to uniform precision}: the unsigned criterion assigns the same bit width
to every unit, so it no longer discriminates between modules at all, and the drift margin in
those cells should be read as a margin over uniform allocation rather than over LieQ. On the
remaining \dlabnonuni\ --- \dlabnonunimodels\ --- LieQ-abs produces a genuinely non-uniform and
usable allocation, and drift is still ahead by \dlabgaplo\ and \dlabgaphi\ points. Those two
cells are the ones that carry the comparison: the criterion has been given its best reading,
the model is alive, the allocation is non-trivial, and measuring the deployed consequence still
allocates better than reading the internal geometry.

The same distinction corrects a rule we initially adopted and then found the ledger
contradicts. It is tempting to discard cells where the same-budget uniform reference has itself
collapsed, on the grounds that everything there has already failed. It has not: uniform is
below the threshold in \dlqunidead\ cells, and in \dlqunirescued\ of them at least one
mixed-precision allocation is nonetheless usable --- on EmbGemma at $B_{2\leftrightarrow3}$
uniform retains \dlqgemmauni\ of full precision while both allocations retain over four fifths
of it. Whether mixed precision rescues a budget that uniform cannot is a question about the
allocation, not a reason to drop the cell.

\paragraph{Do the two criteria agree?} Because everything outside the sensitivity score is
held fixed, any gap between LieQ-adapt and the drift allocation is attributable to the
criterion alone. Over the twelve allocation units the rank correlation between the two
criteria \dagreeverdict: its median is \dagreerho\ (Kendall \dagreetau) across the \dagreen\
measured cells, and in \dagreebottom\ of them the module the drift signal ranks \emph{most}
sensitive falls in the bottom third of the LieQ ordering. Where the two orderings differ on a
band the allocator is deciding, the allocation follows the criterion and the held-out score
follows the allocation, which is why the agreement column and the gain column in
Table~\ref{tab:lieqarms} move together. Two things belong with any reading of that table. The
LieQ diagnostic is not noisy here --- its seed correlation is essentially one, so a low score
would indicate a criterion that ranks differently, not one that ranks randomly --- and LieQ was
designed and validated for generative language models under perplexity, so this is evidence
about how far the criterion transfers to retrieval encoders rather than about the original
result. The mean overlap between the two criteria's three most sensitive modules is
\dagreeoverlap. The reading we support is therefore the narrow one: intrinsic representation
geometry and quantization-induced retrieval drift prioritize substantially different modules,
and this disagreement can materially affect retrieval quality. We deliberately do not claim
that LieQ's criterion misses modules that matter --- that is a causal statement about a
method as described, and these cells are development cells the method was tuned against. The
disagreement metrics, the pre-specified arm, the score orientation and two directional
hypotheses were frozen in advance of opening the three untouched confirmation models, which
are what give this comparison confirmatory status; everything reported here is exploratory.

\begin{figure}[t]\centering
\includegraphics[width=\columnwidth]{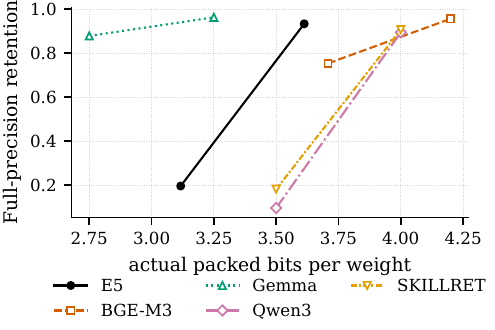}
\caption{Full-precision retention of the drift allocation against the actual packed budget.
Retention is quantized NDCG@10 divided by full-precision NDCG@10 on the same corpus. The same
normalized budget lands in very different regimes on different architectures.}
\label{fig:frontier}
\end{figure}

At $B_{2\leftrightarrow3}$ the drift allocation improves several-fold over uniform
quantization on some models while still retaining a small fraction of full-precision quality.
We report retention alongside absolute NDCG throughout so that this reads as mitigating a
collapse rather than enabling usable inference at that budget. Where a model collapses is
architecture-dependent at a fixed normalized budget; we keep the budget grid common and treat
the collapse frontier as an analysis rather than tuning the operating point per model, since
choosing each model's budget by its held-out score would be selecting the operating point on
downstream supervision.

\subsection{Confirmation on the untouched models}\label{sec:confirm}

\begin{table*}[t]\centering\footnotesize
\begin{tabular}{llcccccccc}
\toprule
Model & Budget & Group & bpw & FP32 & Uniform & LieQ-signed & LieQ-abs & Drift (ours) & Stratum \\
\midrule
mxbai-large & $B_{2\leftrightarrow3}$ & 32 & 3.559 & 38.40 & 5.30 & 27.99 & 27.99 & 28.84 & graded \\
mxbai-large & $B_{3}$ & 32 & 4.056 & 38.40 & 37.41 & 34.83 & 37.41$^{\equiv\mathrm{U}}$ & 36.42 & graded \\
MPNet-base & $B_{2\leftrightarrow3}$ & 64 & 3.117 & 33.57 & 1.04 & 7.61 & 7.61 & 6.09 & dead \\
MPNet-base & $B_{3}$ & 64 & 3.613 & 33.57 & 30.52 & 16.75 & 16.75 & 29.67 & cliff \\
GTE-base & $B_{2\leftrightarrow3}$ & 64 & 3.000 & 36.02 & 1.42 & 1.01 & 0.79 & 2.44 & dead \\
GTE-base & $B_{3}$ & 64 & 3.500 & 36.02 & 28.87 & 1.21 & 27.51 & 27.85 & cliff \\
\bottomrule
\end{tabular}

\caption{\textbf{Confirmation} on the three embedders held untouched until the method,
baselines, budgets, thresholds and hypotheses were frozen and sealed. Held-out NDCG@10, all
arms under the same packed-byte budget, one run per cell under the frozen protocol.
$^{\equiv\mathrm{U}}$marks a LieQ-abs allocation byte-identical to uniform. The stratum is the
pre-registered pairwise classification by each arm's own retention against the $0.5\times$full-precision
usability threshold: \emph{graded} (drift and LieQ-signed both usable), \emph{cliff} (one
usable), \emph{dead} (neither).}
\label{tab:confirm}
\end{table*}

Table~\ref{tab:confirm} holds the cells. Everything in this subsection was fixed before the three models were opened: the balanced
drift of Equation~\ref{eq:drift}, the LieQ-signed and LieQ-abs recipes, the budget grid, the
usability threshold, the random-only predictability pool, the six metrics recorded below and
the two hypotheses. The pre-registration documents and their two amendments are sealed by
hash and the verification script is released. The order of reporting is also pre-registered:
the five development models are exploratory, the three untouched models are the confirmation,
and a pooled figure over eight is secondary and never the first sentence.

\paragraph{The pre-registered hypotheses.} \textbf{H-LieQ} states that at $B_{3}$ the drift
allocation scores above the pre-specified LieQ-signed arm on a majority of the untouched
models and never falls below the usability threshold on any of them. It holds: drift is above
LieQ-signed on \dconfhlieq\ and retains at least three quarters of full precision on all
three. \textbf{H-LieQ-abs}, the secondary hypothesis against the two-sided steelman, holds in
\dconfhabs\ --- on mxbai-large the unsigned criterion selects the uniform plan
byte-for-byte, so that cell is a comparison against uniform and drift scores lower there; on
GTE-base drift scores higher than LieQ-abs by a third of a point, which we do not call a win. Both
hypotheses were stated as directional replications at $n=3$, and that is all they are.

\paragraph{What was not hypothesised, and is the important number.} The pre-registration
compared drift with LieQ, not with uniform precision, because uniform is the trivial
allocation and the development results had it behind drift in every cell
(\ddevuni\ at $B_{3}$). On the untouched models the picture is different. At $B_{3}$ the
drift allocation is numerically lower than same-budget uniform on all three, by
\dconfunipp\ points (\dconfuni\ wins for drift); the cells are graded or cliff-crossing
against LieQ, and the drift plan is never the collapsed one, but the point margin over the
trivial allocation that Table~\ref{tab:baselines} shows on the development models does not
replicate. Whether a one-point difference on \dreevalnq\ queries is a difference at all is
addressed below. At
$B_{2\leftrightarrow3}$ two of the three models are dead for every arm --- uniform, LieQ and
drift all below the threshold --- and on the third, mxbai-large, uniform precision collapses
to \dconfmxbaiuni\ while drift retains \dconfmxbaidrift\ (\dconfmxbailow\ points), against
\dconfmxbaisigned\ for LieQ-signed. Across the six untouched cells the strata are
\dconfgraded\ graded, \dconfcliff\ cliff-crossing and \dconfdead\ dead. The pooled
$n=8$ figures are \dpoolhlieq\ against LieQ-signed and \dpooluni\ against uniform at $B_{3}$;
we give them for completeness and they should not be read in place of the two paragraphs
above.

\paragraph{The remaining pre-declared metrics.} The random-only predictability pool ---
uniformly sampled feasible plans with no allocator neighbourhood --- gives $\rho$ between
\dconfrhomin\ and \dconfrhomax\ on the three models, all above the pre-registered $0.6$ but
with mxbai-large at the floor; the drift signal reproduces under calibration resampling with
seed-pair $\rho \ge \dconfrepro$, Hamming $0$ and byte-identical plans on all three, the same
gate InfoQ-X failed. LieQ-abs coincides with uniform at $B_{3}$ on \dconfabsuni\ of the three
models; the pre-declared \emph{signed-to-abs recovery} --- a signed collapse that the unsigned
criterion lifts back above the threshold --- occurs on \dconfrecovery, where LieQ-signed
scores \dconfgtesigned\ and LieQ-abs \dconfgteabs\ against uniform \dconfgteuni; and the
association between a majority of negative attention $\Delta r$ and a signed collapse holds
on two of the three models. The exception is MPNet-base at $B_{3}$, where only one attention
band is negative yet LieQ-signed retains \dconfmpnetsignedret\ of full precision --- the
usability threshold to three decimals; the cell is classified cliff by the rule as written,
it sits on the boundary, and we do not move the threshold to accommodate it.

\paragraph{Why uniform is not beaten: a three-step decomposition.} The additive allocator
minimises the summed module drift measured on the \emph{calibration} corpus subject to the
byte budget, and at $B_{3}$ the uniform plan is itself feasible, so on that objective the
selected plan is never worse than uniform. For this post-confirmation diagnosis we
independently re-measure the analogous module-wise quantity on the held-out corpus,
$\hat D_{\mathrm{held\text{-}out}}$, the sum of single-module held-out drifts under the
plan; in all three cells the selected plan also has lower $\hat D_{\mathrm{held\text{-}out}}$
than uniform, so the calibration-side preference transfers. That leaves two ways to lose to
uniform on NDCG. Either the whole-model drift $D$
of the selected plan is in fact \emph{higher} than uniform's --- the additive prediction
composed badly --- or $D$ is lower as predicted and retrieval is nevertheless worse, in which
case drift itself does not resolve this comparison. We re-evaluated the untouched $B_{3}$
cells independently --- the ledger's own bit plans, the same held-out protocol, a different
machine and a fresh model load --- recording $\hat D_{\mathrm{held\text{-}out}}$, the realised whole-model drift $D$ and NDCG
for the drift and uniform plans together with a query-level paired bootstrap
(Table~\ref{tab:reeval}). The
re-evaluation reproduces every ledger value to within \dreevalmaxd\ NDCG points and the
sign of every drift-minus-uniform difference. On \dreevalcaseb\ of the \dreevaln\
re-evaluated models the outcome is the second case: the drift plan has lower module-wise
\emph{and} lower realised whole-model drift than uniform, and lower NDCG; on \dreevalcasea\
it is the first. The chain is therefore intact from calibration to deployment ---
calibration drift minimised, held-out module drift lower, whole-model drift lower, on GTE-base
by more than half --- and retrieval quality is nevertheless about a point lower. The loss is
not a failure of additive composition; within the high-quality regime, less representation
movement is not better ranking, shown here on frozen plans rather than as a correlation ---
the fine-selection weakness the development analysis had already measured. The paired 95\% interval for drift minus uniform excludes zero on \dreevalciexcl\ of the
\dreevaln\ re-evaluated models (mxbai-large: \dremxbaidelta\ points,
[\dremxbaicilo, \dremxbaicihi]) and includes it on \dreevalcicross\ (MPNet-base:
\drempnetdelta\ points, [\drempnetcilo, \drempnetcihi]; GTE-base: \dregtedelta\ points,
[\dregtecilo, \dregtecihi]), so
the honest statement is that drift is \emph{numerically lower} by about a point on all three,
that the difference is resolved at \dreevalnq\ queries on some models and not on others, and
that the direction is the same everywhere. One accounting detail should be read alongside
these numbers: the allocator's budget enforcement lands the selected plan within $0.03$ bits
per weight below the budget while uniform sits exactly on it, so the drift allocation uses up
to \dreevalbitsgap\ fewer bits per weight than uniform in these cells; its comparison
therefore does not benefit from exceeding the uniform budget, although fewer bits do not by
themselves imply lower retrieval quality.

\begin{table*}[t]\centering\footnotesize
\begin{tabular}{lccccc}
\toprule
Model & Device & $\max|\Delta|$ vs ledger & $\hat D$ drift / uniform & $D$ drift / uniform & $\Delta$NDCG drift$-$uniform [95\% CI] \\
\midrule
mxbai-large & mps & 0.060 & 0.043 / 0.059 & 0.048 / 0.058 & -1.04 [-1.85, -0.28] \\
MPNet-base & mps & 0.042 & 0.097 / 0.131 & 0.129 / 0.143 & -0.90 [-1.90, 0.18] \\
GTE-base & cuda & 0.000 & 0.131 / 0.289 & 0.145 / 0.326 & -1.01 [-2.71, 0.64] \\
\bottomrule
\end{tabular}

\caption{Independent re-evaluation of the untouched $B_{3}$ cells with the ledger's own bit
plans and held-out protocol on a different machine. $\max|\Delta|$ is the largest absolute
difference from the ledger over the full-precision, uniform, drift and searched arms.
$\hat D_{\mathrm{held\text{-}out}}$ is the sum of single-module held-out drifts under the plan
(the allocator itself minimised the calibration-corpus analogue), $D$ the measured
whole-model balanced drift. The last column is the query-level paired bootstrap of drift
minus uniform NDCG@10 (5{,}000 resamples, \dreevalnq\ queries; percentile interval).}
\label{tab:reeval}
\end{table*}

\paragraph{What we conclude.} The confirmation supports the narrow, pre-registered claim ---
against the prior geometry criterion, the label-free drift signal allocates better and
never collapses --- and it does not support the broader reading a development table invites,
that drift beats the trivial allocation by a point or two whenever the model is usable. On
these three models, at the budget where uniform precision is usable, drift is numerically
lower by about a point, with query-level intervals that exclude zero on some models and not
others; in the one untouched
cell where uniform collapses but a tested mixed-precision allocation remains usable, drift
remains usable as well (\dconfmxbaidrift\ against \dconfmxbaiuni\ for uniform), although
LieQ-signed also survives at \dconfmxbaisigned. The seeded local search arm, which the
development results told us not to adopt, finds a plan \dconfgtesearch\ points above uniform
on GTE-base at $B_{3}$, so headroom over uniform exists there and the additive drift
allocation does not find it. We keep the development-stage decision rather than revisiting it
after seeing the confirmation, and record the headroom as the next question. The searched
figures on the untouched models are descriptive only: the pre-registered adoption decision
was closed on the ten development cells and is not reopened here.

\subsection{Cross-corpus transport}\label{sec:transfer}

\begin{table*}[t]\centering\footnotesize
\begin{tabular}{lcccccc}
\toprule
Model & \multicolumn{2}{c}{Probe$\to$held-out} & \multicolumn{4}{c}{Four-probe stability} \\
\cmidrule(lr){2-3}\cmidrule(lr){4-7}
 & $\rho$ & cells & $\rho$ & identical plans & median Hamming & NDCG spread \\
\midrule
E5-base & 0.992 & 2 & 0.993 & 12/12 & 0.0 & 0.00 \\
BGE-M3 & 0.996 & 2 & 0.997 & 8/12 & 0.0 & 0.05 \\
EmbGemma & 0.986 & 2 & 0.989 & 7/12 & 0.0 & 0.37 \\
Qwen3-0.6B & 0.991 & 2 & 0.993 & 9/12 & 0.0 & 11.43 \\
SKILLRET & 0.994 & 2 & 0.996 & 12/12 & 0.0 & 0.00 \\
\bottomrule
\end{tabular}

\caption{Transport of module sensitivity across retrieval domains. The left block correlates
module drift measured on the probe corpus against the same cells measured on the held-out
corpus. The right block varies the probe over four corpora and reports the mean pairwise
sensitivity correlation, how many probe pairs yield a byte-identical bit plan, the median
Hamming distance between plans, and the spread of held-out NDCG across probes. Probe bank and
evaluation bank are disjoint.}
\label{tab:transfer}
\end{table*}

\begin{table*}[t]\centering\footnotesize
\begin{tabular}{lccccc}
\toprule
 & & \multicolumn{2}{c}{$B_{2\leftrightarrow3}$} & \multicolumn{2}{c}{$B_{3}$} \\
\cmidrule(lr){3-4}\cmidrule(lr){5-6}
Model & $\rho$ & Identical & Max $\Delta$NDCG & Identical & Max $\Delta$NDCG \\
\midrule
E5-base & 0.992 & 100.0\%$^{\dagger}$ & 0.000$^{\dagger}$ & 100.0\% & 0.000 \\
BGE-M3 & 0.998 & 33.3\% & 0.053 & 100.0\% & 0.000 \\
EmbGemma & 0.987 & 100.0\% & 0.000 & 16.7\% & 0.372 \\
Qwen3-0.6B & 0.991 & 50.0\%$^{\dagger}$ & 11.425$^{\dagger}$ & 100.0\% & 0.000 \\
SKILLRET & 0.994 & 100.0\%$^{\dagger}$ & 0.000$^{\dagger}$ & 100.0\% & 0.000 \\
\midrule
All cells ($n=10$) & 0.992 & \multicolumn{2}{c}{80.0\% identical} & \multicolumn{2}{c}{max 11.425} \\
Non-collapsed ($n=7$) & 0.992 & \multicolumn{2}{c}{78.6\% identical} & \multicolumn{2}{c}{max 0.372} \\
\bottomrule
\end{tabular}

\caption{Four-probe domain stability. Module sensitivity is measured independently on SciFact,
FiQA, ArguAna and NFCorpus; every resulting plan is evaluated on SciDocs, which appears in no
probe. $\rho$ is the median pairwise Spearman of the $12 \times 3$ module-by-bit sensitivities
over the six probe pairs. Identical is the fraction of probe pairs selecting a byte-identical
bit plan at that budget, and Max $\Delta$NDCG the largest held-out disagreement any pair
produces. $^{\dagger}$marks cells the budget collapses (best probe retains under half of
full-precision). Those cells are reported separately rather than dismissed: all of their
resulting systems lie in the collapsed regime, so a plan disagreement there is not comparable
to one between usable systems, but it is not evidence that the signal is stable either --- the
Qwen3-0.6B cell shows the signal-to-decision chain becoming \emph{more} consequential past the
cliff. The macro row is computed over the non-collapsed cells and the all-cell macro is given
in the text.}
\label{tab:fourprobe}
\end{table*}

Relevance-derived sensitivity was, in our earlier measurements, strongly dependent on which
corpus produced it. Drift is much less so: module sensitivity measured on one retrieval corpus
predicts the same quantity on another at \dtransfer, taken over the \dtransfern\ development models.
Table~\ref{tab:transfer} separates the three things that could break --- the sensitivity
ordering, the allocation decision, and the downstream score --- because they can disagree.
Matching sensitivities with differing plans indicates an allocator threshold effect rather
than an unstable signal; differing plans with matching NDCG indicates a flat optimum.

Table~\ref{tab:fourprobe} varies the probe over four corpora --- SciFact, FiQA, ArguAna and
NFCorpus --- and evaluates every resulting plan on a fifth, SciDocs, which appears in no
probe. Across the five development models the sensitivity ordering barely moves: the median
pairwise Spearman over all module-by-bit cells is \dfpmacro. Over all \dfpalln\ cells,
\dfpallident\ of probe pairs select a byte-identical bit plan, with a worst-case held-out
disagreement of \dfpallspread\,points; restricted to the \dfpn\ cells whose budget leaves a
usable model, \dfpident\ agree exactly and the worst disagreement falls to
\dfpspread\,points. The stratification is by a criterion fixed earlier for the compression
frontier --- a cell is collapsed when the best probe retains under half of full-precision
quality --- and not by the domain-stability numbers themselves; both strata are reported so
that the split can be checked. The one cell where plans differ appreciably is EmbeddingGemma at $B_{3}$,
where five of six pairs disagree yet the score spread is \dfpspread\,points --- an allocator
threshold effect on a flat optimum, which is the benign shape. We read this as evidence that
the sensitivity ordering does not depend on which retrieval domain produced it, within the
budgets where the quantized model is usable at all. Stated as one sentence, what the three
columns show is that \emph{sensitivity transport is stronger than plan identity, and plan
disagreement need not imply retrieval disagreement}: the ordering survives the change of probe
almost intact, the allocator's threshold turns small ordering differences into different plans
more often than the ordering itself moves, and where the optimum is flat those different plans
land on nearly the same score.

\dfpcollapse\ cells fall in the collapsed stratum and are marked in the table. Qwen3-0.6B at $B_{2\leftrightarrow3}$ is the instructive
one: probe pairs disagree by \dfpqwen\,points of NDCG@10, which looks like a large domain
effect until one notes that the arms being compared score \dfpqwenlo\ and \dfpqwenhi\ against
a full-precision \dfpqwenfp. Both are unusable, so the disagreement is not a difference between
two deployable systems. It is still informative: it says that domain robustness itself can
break down once a budget has pushed a model past its utility cliff, which is why the collapsed
stratum is reported rather than discarded. The claim the paper makes is the stratified one ---
transport is stable in the usable compression regime, and aggressive compression can amplify
probe dependence after the model has already stopped being usable.

This is the measurement whose absence removed ``domain-stable'' from the title. It is now
present for the five development models, and the claim is still held back: the four-probe
measurement was not part of the frozen confirmation matrix and has not been run on the three
untouched models, and a property demonstrated on the models used to develop the method is not
yet a property of the method.

\section{Module drift preserves configuration ordering, not
         magnitude}\label{sec:composition}

\begin{figure*}[t]\centering
\includegraphics[width=\textwidth]{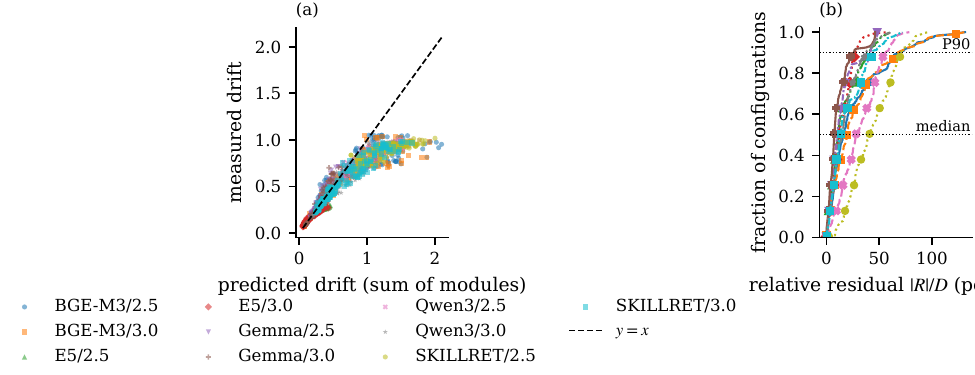}
\caption{(a) Predicted drift $\sum_i d_{i,b_i}$, the sum of single-module drifts, against the
drift measured by quantizing the whole configuration at once; the dashed line is $y=x$.
(b) Empirical CDF of the relative residual $|R|/D_{\mathrm{whole}}$ for the same
configurations. The cloud in (a) sits systematically below $y=x$ --- the compressive bias
quantified in Table~\ref{tab:composition} --- and (b) shows how far the disagreement reaches:
a tenth of configurations are off by more than a third. Panel (a) read alone invites
``near-additive''; (b) is why the text below claims ordering rather than magnitude.
Module drifts and configuration drifts are measured on the same corpus, so the spread here is
composition error and not cross-corpus transfer.}
\label{fig:additivity}
\end{figure*}

\begin{table*}[t]\centering\footnotesize
\begin{tabular}{lcccccc}
\toprule
 & \multicolumn{3}{c}{Pairwise inversion rate} & \multicolumn{3}{c}{Calibration $D=\alpha+\beta\hat{D}$} \\
\cmidrule(lr){2-4}\cmidrule(lr){5-7}
Cell & All (\%) & Top 25\% & Seed nbhd. & $\beta$ & $R^2$ & nMAE/IQR \\
\midrule
E5-base $B_{2\leftrightarrow3}$ & 10.7\% & 18.5\% & 13.3\% & 0.34 & 0.84 & 1.02 \\
E5-base $B_{3}$ & 6.4\% & 16.7\% & 12.0\% & 0.66 & 0.93 & 0.23 \\
BGE-M3 $B_{2\leftrightarrow3}$ & 12.4\% & 23.1\% & 15.5\% & 0.46 & 0.77 & 0.45 \\
BGE-M3 $B_{3}$ & 12.4\% & 6.2\% & 6.7\% & 0.51 & 0.78 & 0.47 \\
EmbGemma $B_{2\leftrightarrow3}$ & 5.2\% & 11.4\% & 9.9\% & 0.65 & 0.90 & 0.20 \\
EmbGemma $B_{3}$ & 4.5\% & 4.5\% & 5.1\% & 0.77 & 0.89 & 0.13 \\
Qwen3-0.6B $B_{2\leftrightarrow3}$ & 16.8\% & 29.8\% & 18.2\% & 0.36 & 0.76 & 1.33 \\
Qwen3-0.6B $B_{3}$ & 7.5\% & 4.9\% & 12.2\% & 0.69 & 0.92 & 0.27 \\
SKILLRET $B_{2\leftrightarrow3}$ & 16.9\% & 33.7\% & 20.0\% & 0.26 & 0.72 & 2.40 \\
SKILLRET $B_{3}$ & 6.6\% & 4.8\% & 8.8\% & 0.60 & 0.92 & 0.35 \\
\midrule
Macro ($n=10$) & 9.9\% & 15.3\% & 12.2\% & 0.53 & 0.84 & 0.69 \\
\bottomrule
\end{tabular}

\caption{How module drift composes. The inversion rate is
$\Pr[(\hat{D}_A - \hat{D}_B)(D_A - D_B) < 0]$ over configuration pairs, restricted to three
regions: the whole pool, the top quartile by true held-out NDCG, and the budget-preserving
neighbourhood of the additive plan --- the region a local search actually explores. The
calibration columns fit $D = \alpha + \beta\hat{D}$; nMAE/IQR is the mean absolute residual
divided by the interquartile range of measured drift, a scale-free magnitude figure that does
not inherit the small-denominator inflation of $|R|/D$.}
\label{tab:composition}
\end{table*}

Summing single-module drifts predicts the drift of the whole configuration with a rank
correlation of \daddspearman\ but a median relative residual of \daddresid\ and a P90 near
\daddpninety\ (Table~\ref{tab:additivity}). That relative figure has a denominator artifact:
configurations with small measured drift inflate $|R|/D$. Table~\ref{tab:composition}
therefore reports the same disagreement on scale-free axes, and the picture holds --- the mean
absolute residual is \dnmaeiqr\ of the interquartile range of measured drift.

The regression makes the shape of the disagreement precise. Fitting
$D = \alpha + \beta\hat{D} + \varepsilon$ gives a \emph{compressive} trend: $\beta$ lies in
$[\dbetalo, \dbetahi]$ across cells with $R^2 \approx \dcalibrsq$. Composition error is
therefore systematic rather than random --- the sum consistently overstates how much a
configuration will move, increasingly so for aggressive configurations.

It is worth being precise about what this does and does not explain, because the obvious
reading is wrong. The \emph{fitted mean trend} $\alpha + \beta\hat{D}$ is monotone, which is
why the global ordering is largely preserved; the \emph{relation itself} is not monotone,
because $\varepsilon$ can and does reverse pairs. Measured directly, \dinvall\ of
configuration pairs are ordered wrongly by the module sum overall, rising to
\dinvtoptwentyfive\ within the top quartile by retrieval quality and \dinvseed\ inside the
additive plan's own neighbourhood. A monotone fit does not make the ordering exact, and the
two facts are only consistent when stated together: \emph{the fitted mean trend is compressive
and monotone, explaining why global ordering is largely preserved; non-negligible residuals
nevertheless produce pairwise inversions, particularly among high-quality neighbouring
configurations.}

The residual is largest exactly where the damage is largest: $|R|$ correlates with the number
of 2-bit modules and with predicted drift magnitude, and negatively with nominal bits per
weight. The summary we take forward is that \emph{interactions are substantial in magnitude
but usually insufficient to disrupt the global configuration ordering; they can still matter
within narrow high-quality neighbourhoods.} We use \emph{rank-compositional} for this property
rather than ``near-additive'', which overstates what the magnitude evidence supports.
\S\ref{sec:analysis} shows the algorithmic consequence.

We report $\beta$ and $R^2$ as descriptive statistics and deliberately do not attach a
significance test to $\beta \ne 1$. The configurations in each pool are not an i.i.d.\ sample:
they are structurally generated and share module assignments heavily, so the sampling
population an OLS $t$-statistic would refer to is not well defined here. Establishing
significance would require first fixing a configuration sampling design and then
bootstrapping over it.

\section{Analysis: why is the allocator this simple?}\label{sec:analysis}

Appendix~\ref{app:nulls} asks the prior question --- whether the gain comes from the signal
at all rather than from the allocator's structure --- with permutation and random-feasible
nulls; the answer is yes on every development cell, and the subsections below take that as
given.

\subsection{Does relevance supervision help?}\label{sec:labels}

Our first design derived sensitivity from measured NDCG damage per module. On two models it
was the weakest of seven signals at predicting held-out damage, its allocation lost to drift,
and it did not clear a null built by shuffling the relevance judgements fifty times and
running the allocator to completion on each. Controlling for drift, the partial correlation
between label-derived sensitivity and true damage was approximately zero. We found one cell
where labels produced the better allocation and no pre-allocation criterion that identifies
such cells in advance. The accurate statement is therefore not that labels carry no
information, but that \emph{relevance-derived sensitivity provided no consistent incremental
predictive value beyond drift under our tested allocator and calibration protocol}; labels are
reported as an ablation and are absent from the method.

\subsection{Does local configuration search help?}\label{sec:search}

\begin{figure}[t]\centering
\includegraphics[width=\columnwidth]{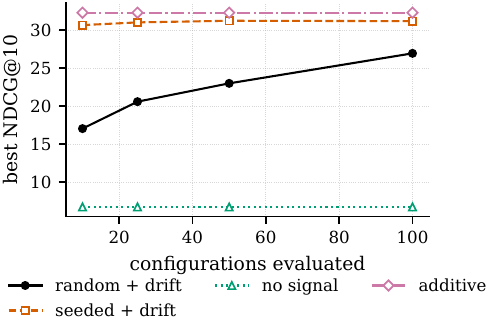}
\caption{Budget-matched candidate selection. Each curve reports the best held-out NDCG@10
reached after evaluating $k$ configurations, selecting by label-free whole-model drift alone;
NDCG is never consulted during selection. ``no signal'' picks uniformly at random from the
same candidates. ``additive'' is the drift allocation, which costs 36 single-module encodings
rather than $k$ configuration evaluations. Curves are single trajectories; seed variance is
not measured (\S\ref{sec:limits}).}
\label{fig:efficiency}
\end{figure}

\begin{table*}[t]\centering\footnotesize
\begin{tabular}{llcccccc}
\toprule
Budget & Model & Uniform & Additive & Rand+drift & Seed+drift & $\Delta_{\mathrm{search}}$ & $\Delta_{\mathrm{seed-rand}}$ \\
\midrule
$B_{2\leftrightarrow3}$ & E5-base & 1.63 & 6.79 & 4.20 & 7.18 & 0.390 & 2.985 \\
 & BGE-M3 & 0.97 & 23.75 & 18.98 & 25.10 & 1.346 & 6.118 \\
 & EmbGemma & 1.49 & 33.86 & 30.20 & 33.34 & -0.524 & 3.138 \\
 & Qwen3-0.6B & 0.97 & 3.38 & 8.68 & 7.96 & 4.581 & -0.724 \\
 & SKILLRET & 1.52 & 6.15 & 8.60 & 14.74 & 8.582 & 6.135 \\
\midrule
$B_{3}$ & E5-base & 31.24 & 32.29 & 26.94 & 31.18 & -1.111 & 4.242 \\
 & BGE-M3 & 27.67 & 30.15 & 28.48 & 30.37 & 0.220 & 1.886 \\
 & EmbGemma & 36.13 & 37.12 & 35.07 & 36.88 & -0.245 & 1.813 \\
 & Qwen3-0.6B & 31.00 & 31.31 & 24.76 & 31.76 & 0.456 & 7.003 \\
 & SKILLRET & 29.01 & 30.62 & 26.03 & 30.89 & 0.269 & 4.858 \\
\bottomrule
\end{tabular}

\caption{Held-out NDCG@10 of three arms at one evaluation budget of \dnrandom\ configurations.
$\Delta_{\mathrm{search}}$ is seeded search minus the additive plan;
$\Delta_{\mathrm{seed-rand}}$ is seeded search minus random search. Selection within both
search arms uses drift only; NDCG is never consulted during selection.}
\label{tab:arms}
\end{table*}

Seeding a configuration search from the drift allocation beats seeding it randomly in nine of
the ten cells (Table~\ref{tab:arms}), so module drift is a useful proposal distribution; the
exception is Qwen3-0.6B at $B_{2\leftrightarrow3}$, where random candidates happened to
include a better plan than the seeded neighbourhood contained.

Against the additive plan itself the effect is strongly \emph{heterogeneous} rather than
uniformly absent, and this is the finding that changed most as the matrix filled in. With four
cells measured the median $\Delta_{\mathrm{search}}$ was negative and the natural reading was
that search does not help. With nine it is positive, and two cells improve by several NDCG
points. Table~\ref{tab:headroom} gives the first half of the reason: in part of the grid
\emph{no} budget-preserving neighbour of the additive plan is better, so the question of
selector quality does not arise there --- these are the cells whose feasible neighbourhood is
small because the budget coincides with a uniform grid point.

The second half is that the cells where search wins are exactly the cells where the additive
surrogate composes worst. Across the measured cells, $\Delta_{\mathrm{search}}$ correlates
with the composition error of \S\ref{sec:composition} at $r = \dheadroomnmae$ against
nMAE/IQR and $r = \dheadroombeta$ against $\beta$: the more compressive and the noisier the
composition, the further the additive plan sits from its neighbourhood optimum and the more a
search recovers. This is what the compressive-affine account predicts, and, exploratorily,
it means the useful statement is not ``search does not help'' but \emph{search gain
concentrates where the additive surrogate composes poorly}; the relationship is a post-hoc
observation on the ten development cells and was not tested by a sealed prediction on the
untouched models.

The heterogeneity aligns with the budget. At the stress point
$B_{2\leftrightarrow3}$ the median gain is \dsearchlow\,pp with \dsearchlown\ of five cells
improving; at the primary operating point $B_{3}$ it is \dsearchmid\,pp with
\dsearchmidn\ of five. Aggressive budgets are where the additive surrogate composes worst
(Table~\ref{tab:composition}) and where a search therefore has something to find; in the
development matrix the $B_{3}$ gains were typically smaller, because uniform quantization
already retains most of full-precision quality there. That is a statement about these five
models, not a law: on the untouched GTE-base at $B_{3}$ the same seeded search finds
\dconfgtesearch\ points over uniform (\S\ref{sec:confirm}).

It matters to keep two things apart here. Search \emph{failed our pre-registered criterion}
for inclusion in the default method, despite improving \dsearchimproved\ of \dsearchcells\ cells; post-hoc stratification
revealed strong budget-dependent heterogeneity, with substantially larger gains in the
aggressive low-bit regime. The criterion was fixed before any cell was observed and is
applied as written: \dsearchverdict. That the rule excludes an intervention which helped in
most cells is a property of requiring both a minimum median effect and a worst-case floor,
and we report it rather than revising the rule after the fact.

Two cautions on the relationship itself. It is a post-hoc observation on \dheadroomn\ cells,
it was not pre-registered, and $\Delta_{\mathrm{search}}$ and the neighbourhood inversion rate
are computed from the same seeded arm, so they are not independent. We therefore treat it as
an exploratory hypothesis. No such prediction was sealed before the untouched models were
run, so this hypothesis was not tested on them and remains exploratory; the search arm itself
did run there as part of the frozen protocol and is reported in \S\ref{sec:confirm}.

\paragraph{Label-free is not the same as cheap.} A natural next step is to use the
composition error as a gate: run the expensive search only where the cheap allocator is
predicted to be insufficient. Stated naively this does not work, and the reason is worth
recording. Estimating $\beta$ or nMAE/IQR from the full pool requires evaluating one to two
hundred whole configurations, which is most of the cost of the search it is supposed to
authorise. A usable gate has to be built from a handful of pre-specified sentinel
configurations near the additive plan, and whether a handful suffices is an empirical
question about estimator variance, not an assumption. \S\ref{sec:limits} records the sample
size and the statistic that survive that test, and the escalation design they imply.

We keep this gate out of the method proper, for a reason that is easy to lose track of. The
allocator of \S\ref{sec:method} uses no relevance judgements at any stage and is label-free
without qualification. A gate whose threshold is calibrated against measured
$\Delta_{\mathrm{search}}$ has consumed relevance labels \emph{during development}, even
though applying it at deployment needs none. The honest description of such a gate is
\emph{label-free at deployment}, not label-free, and mixing the two would let a reader
attribute the stronger property to the whole method. Composition-gated search is therefore
reported as an optional extension whose confirmation is still outstanding.

\section{Related work}\label{sec:related}

\paragraph{Uniform and mixed-precision PTQ.} Post-training quantization is surveyed
in~\cite{gholami2021survey}. Its reconstruction branch optimizes rounding or weights against a
layer- or block-wise output objective on a small calibration set, from
AdaRound~\cite{nagel2020adaround} and BRECQ~\cite{li2021brecq} to Optimal Brain
Compression~\cite{frantar2022obc}. For large transformers, weight-only PTQ is dominated by
GPTQ~\cite{frantar2023gptq} and activation-aware scaling as in AWQ~\cite{lin2024awq}; LLM
quantization more broadly adds outlier-, sensitivity- and incoherence-aware schemes such as LLM.int8()~\cite{dettmers2022llmint8},
SmoothQuant~\cite{xiao2023smoothquant}, ZeroQuant~\cite{yao2022zeroquant},
SpQR~\cite{dettmers2024spqr}, SqueezeLLM~\cite{kim2024squeezellm},
OmniQuant~\cite{shao2024omniquant} and QuIP~\cite{chee2023quip}. These methods chiefly improve how a
tensor is quantized at a given precision; they are orthogonal to how precision is
\emph{distributed}, and our quantizer is deliberately the plain round-to-nearest baseline so that
the allocation signal is the only variable. Mixed-precision allocation adds a per-layer
sensitivity score, classically the Hessian trace of HAWQ~\cite{dong2019hawq,dong2020hawqv2} and
its integer-only successor HAWQ-V3~\cite{yao2021hawqv3}, or a search policy with hardware
feedback as in HAQ~\cite{wang2019haq}.

\paragraph{Quantizing transformer encoders.} Encoder-only transformers were quantized well
before generative LLMs: Q8BERT~\cite{zafrir2019q8bert} applied byte-width quantization-aware
training, Q-BERT~\cite{shen2020qbert} assigned group-wise mixed precision from Hessian spectra,
and I-BERT~\cite{kim2021ibert} made inference integer-only. All three are evaluated on
classification and span-extraction benchmarks, where the task head sits on top of the encoder
and absorbs part of the perturbation. An embedder has no such head: its pooled output is the
product, which is why we observe it directly.

\paragraph{Geometry- and information-based allocation.} LieQ~\cite{lieq2025} allocates bits in
sub-8B language models from layer-wise diagnostics. It motivates the criterion with a
perplexity-drop measurement of functional saliency, but the deployed score is purely
geometric and needs no perplexity probing: for each projection it compares the spectral
entropy of the layer's output against the same input passed through an untrained weight
matrix of the same initialization distribution, and reads the difference as
training-induced energy concentration. Layers whose spectra training concentrated most are
treated as functionally irreplaceable and given the higher precision. This is the sharpest
available contrast with our signal, because it is measured entirely on the full-precision
model: nothing is quantized before the score is read. We quantize first and observe what
moves in the output the retriever consumes. \S\ref{sec:against} puts the two criteria head to
head under an identical allocator and budget, so the comparison isolates the sensitivity
criterion rather than the surrounding pipeline.

\paragraph{Intervention-based allocation.} InfoQ~\cite{akbulut2026infoq} is the closest prior
work in mechanism. It quantizes each layer at each candidate bit width and scores it by the
change in mutual information among the subsequent layers, then solves an integer program.
Our allocator is the same program; our score is a different observable in a different setting
--- final embedding displacement rather than internal information flow, dense retrieval rather
than image classification, and cross-domain transport of the score rather than a single
calibration set. We do not claim intervention-based sensitivity as novel, and we regard an
InfoQ-style adaptation as the baseline this paper most needs. One structural obstacle should
be stated plainly rather than papered over. InfoQ's full criterion relies on class-labelled
task information and accuracy-based observer selection. Dense retrieval has pairwise
relevance rather than a per-example class target, so there is no canonical one-to-one port of
its $I(L;Y)$ branch. We therefore evaluate the directly transferable information-flow
component, and call the result an \emph{InfoQ-X adaptation} rather than a reproduction.

\paragraph{Interaction-aware allocation.} CoopQ~\cite{zhao2026coopq} models inter-layer
dependence explicitly, estimating Shapley values for layer sensitivities and their
interactions and solving a binary quadratic program over 2/4-bit assignments.
MixQuant~\cite{mixquant2026} makes the related point that a layer's sensitivity depends on the
bit widths of its upstream layers, and marginalizes each layer's distortion over randomly
quantized upstream configurations. Our composition measurement in \S\ref{sec:composition} is a
contrastive observation against this line rather than a refutation: in the feasible region for
text embedders the residual interaction is large in magnitude and systematically compressive,
yet rarely reorders configurations at the global scale, while remaining large enough to matter
inside a narrow neighbourhood. We have not run an interaction-aware allocator head to head,
and this measurement does not establish that we would beat one.

\paragraph{Compression in dense retrieval.} Most work labelled ``quantization'' in dense
retrieval targets the output vectors or the search index rather than the encoder weights: JPQ
trains the query encoder jointly with a product-quantized index under a ranking
loss~\cite{zhan2021jpq}, and RepCONC learns discrete document representations via constrained
clustering~\cite{zhan2022repconc}. Index-side compression in libraries such as
FAISS~\cite{johnson2017faiss} and truncatable representations such as Matryoshka
embeddings~\cite{kusupati2022matryoshka} shrink the stored vectors, and late-interaction models
such as ColBERT~\cite{khattab2020colbert} move cost from the encoder to the index. Those methods
compress what the encoder \emph{produces};
the setting here is weight-level mixed precision for the encoder itself, which is complementary
and composes with them.

\paragraph{Prior work in one sentence.} Existing methods predominantly estimate sensitivity
from intrinsic geometry, reconstruction error, or interaction-aware proxies, and where they do
intervene they read an internal quantity; we measure the behavioural consequence of a
quantization intervention on the embedder's \emph{output} representation, which is the object
the retrieval metric is computed from.

\section{Limitations}\label{sec:limits}

The most consequential limitations, in order:

\begin{itemize}
  \item \textbf{Published baselines require adaptation, and one of them we could not
        accept.} No published mixed-precision allocator is native to this setting: the
        candidates were designed for generative language models scored by perplexity, over
        per-layer units, with a per-example class target available. We therefore report a
        controlled LieQ adaptation, and we attempted an InfoQ-X adaptation under an estimator
        stability protocol fixed in advance. InfoQ's own estimator is not substitutable ---
        it estimates sliced mutual information with a KSG estimator over random
        one-dimensional projections, so a cheaper information proxy would be a weaker method
        wearing InfoQ's name --- and its $I(L;Y)$ branch has no canonical port to a setting
        without per-example class labels. The adaptation reached the original work's
        \dinfoqxceil-sample scale and did not satisfy the pre-specified criterion, so
        \textbf{it was rejected before any retrieval metric was computed}, not left unrun
        (Appendix~\ref{app:infoqx}). What this costs the paper is concrete: the reader gets a
        controlled comparison against one prior criterion rather than two, and the second
        comparison is reported as a null on estimator reproducibility rather than on
        retrieval quality. InfoQ's downstream quantization-aware training stage is omitted in
        any case, since every arm here is evaluated under identical PTQ-only quantization.
        Our own static-geometry operationalization is not a published system, and its
        ``post-hoc'' column selects on the held-out set, so it is an upper bound for that
        family rather than a competitor.
  \item \textbf{LieQ-adapt is a controlled adaptation, not LieQ's deployment format.} LieQ
        allocates per layer and keeps precision uniform inside a layer so that standard kernels
        apply. We transfer its geometry criterion into our common twelve-unit
        packed-byte framework, which is what makes the sensitivity comparison controlled, but
        it is not LieQ's native setting. Table~\ref{tab:lieqoracle} reports its own two-level
        allocation rule separately for that reason.
  \item \textbf{At the main budget, drift does not beat uniform on the untouched models.}
        The pre-registered hypotheses concerned LieQ and hold; the comparison a reader most
        wants, against the trivial allocation, goes the other way by about a point on all
        three (\S\ref{sec:confirm}). The development table should be read with that
        confirmation next to it, not instead of it.
  \item \textbf{Domain stability is shown only on the development models.} The four-probe
        measurement is now complete for all five development models
        (Table~\ref{tab:fourprobe}) and is strong: \dfpident\ of probe pairs agree on a
        byte-identical plan and the worst held-out disagreement is \dfpspread\,points. It is
        nevertheless measured on the models the method was developed against; it was not part
        of the frozen confirmation matrix and has not been run on the untouched models, so
        ``domain-stable'' stays out of the title.
  \item \textbf{The predictability pool is not fully independent of the signal.} A quarter of
        each pool is perturbations of allocator outputs (\S\ref{sec:setup}), so the pooled
        $\rho$ is measured partly on the drift allocation's own neighbourhood. We therefore
        re-measured it on a drift-independent pool of uniformly sampled feasible plans on all
        five development models: the purely random pools give $\rho$ between \drandpuremin\
        and \drandpuremax\ (median \drandpure) against a mixed-pool median of \drandmixed,
        so on the development models the headline correlation does not rest on the
        allocator's own neighbourhood. On the three untouched models the random-only pool is the primary form
        and gives $\rho$ between \dconfrhomin\ and \dconfrhomax\ --- above the pre-registered
        $0.6$ on all three, and weaker than the development models on two of them.
  \item \textbf{An earlier version of Equation~\ref{eq:drift} pooled queries and documents by count.} With
        $|D|{:}|Q| = 10{:}1$ the pooled statistic is predominantly document drift, while the
        margin bound of Proposition~\ref{prop:bound} charges the two sides equally. This was
        resolved before the untouched models were opened: the balanced, pooled and document-only
        variants select the same plan in all ten development cells while query-only differs in
        two (Appendix~\ref{app:pooling}); the balanced form was
        frozen as Equation~\ref{eq:drift}, and the development ledger measured under the
        pooled form is retained with that audit stated in \S\ref{sec:method}.
  \item \textbf{The composition-risk gate is developed but remains unconfirmed.} Resampling the
        existing pools shows that a sentinel sample of twelve configurations recovers the
        cell ordering of the full-pool slope ($\rho = 0.88$ against the full-pool $\beta$,
        and $\rho = 0.87$ against $\Delta_{\mathrm{search}}$), while every residual-based
        statistic we tried collapses at that sample size ($\rho$ between $-0.17$ and
        $+0.08$). The escalation rule is therefore a threshold on $\beta$ alone, measured on
        the sentinel scale --- sentinel slopes are biased upward relative to the full pool,
        so a threshold fitted on full-pool values would pass everything. The threshold is now fitted:
        on \dctcells\ development endpoints the two cells that gained from search score
        \dcthigh\ and the two that did not score \dctnon, separating cleanly at
        \dctthresh\ with \dctgray\ ambiguous cells. Gating on it would have cost nothing on
        those endpoints, against \dctregretalways\ points for always searching and
        \dctregretnever\ for never searching. \textbf{This is a supervised calibration, not a
        label-free threshold} --- the labels come from held-out $\Delta_{\mathrm{search}}$ ---
        and four endpoints cannot support a threshold. It stays an optional extension; no
        prediction was sealed before the untouched models were run, so the gate was not
        exercised on them, and it is reported as exploratory development work rather than as
        part of the method.
  \item \textbf{Intervals only where the comparison turns on them.} Query-level paired
        bootstrap intervals are given for the drift-versus-uniform comparison on the untouched
        $B_{3}$ cells (Table~\ref{tab:reeval}), and they do not all agree on significance; the remaining held-out
        NDCG figures are point estimates, and the search curves in Figure~\ref{fig:efficiency}
        are single trajectories rather than a median over seeds.
  \item \textbf{Budget-grid coincidence.} Four of the five frozen development budgets coincide
        with that model's uniform 3-bit packed size, which constrains the feasible set and
        weakens the permutation nulls computed there. Nulls are conditional on model and
        budget and must not be compared across either.
  \item \textbf{Deployment format.} Per-band precision on the embedding table assumes a
        row-blocked packing format; we account for its metadata in the budget but report no
        kernel-level latency or throughput.
  \item The composition result establishes rank-compositionality only, and does not imply that
        the additive allocation is a local optimum --- a systematic compressive residual can
        still reorder one.
\end{itemize}

\section{Conclusion}\label{sec:conclusion}

Quantization-induced output drift is a stable, retrieval-facing coarse sensitivity signal: it
is deterministic given a calibration corpus, it reproduces under calibration resampling on
every model we measured, on the development models it transports across retrieval domains in
the usable regime, and it orders sampled allocations against held-out quality well over the
broad sampled configuration pool. Under a
pre-registered protocol on three embedders held untouched until the method was frozen, the
allocation it produces avoids the severe misallocations that a transferred signed-geometry
criterion makes, and at the main $B_{3}$ operating point it never collapses. The same protocol did not reproduce the development
advantage over uniform precision: at the healthy 3-bit operating point drift is numerically
lower on all three models, and the decomposition shows it is lower having reduced module,
held-out and whole-model drift exactly as designed. Drift is much better at identifying dangerous
allocations than at discriminating among already-good ones. The result
suggests that the next problem is not to find another global sensitivity proxy, but to decide
when coarse representation preservation is sufficient and when fine-grained, ranking-aware or
configuration-level refinement is required.

\section*{Reproducibility}

All reported values, tables and plots are generated from a common experiment ledger with
snapshot validation; no number is typed into the source. Every required cell is
ledger-backed, and a missing required cell fails the build rather than rendering as a blank.

\bibliographystyle{plain}
\bibliography{references}

\appendix

\section{Proof of Proposition~\ref{prop:bound}}\label{app:proof}

All embeddings are $\ell_2$-normalized, so $\|E_0(x)\| = \|E_Q(x)\| = 1$ and
\[
  \|E_Q(x) - E_0(x)\|_2^2 = 2 - 2\langle E_Q(x), E_0(x)\rangle = 2\delta(x).
\]
Writing $q_0 = E_0(q)$, $q_1 = E_Q(q)$ and likewise for $d$, add and subtract $\langle q_1,
d_0\rangle$ and apply Cauchy--Schwarz with the unit norms:
\[
\begin{aligned}
  &|\langle q_0, d_0\rangle - \langle q_1, d_1\rangle| \\
  &\quad\le |\langle q_0 - q_1, d_0\rangle| + |\langle q_1, d_0 - d_1\rangle| \\
  &\quad\le \|q_0 - q_1\| + \|d_0 - d_1\| \\
  &\quad= \sqrt{2\delta(q)} + \sqrt{2\delta(d)} .
\end{aligned}
\]
The margin bound follows by applying Equation~\ref{eq:score-bound} to $(q,p)$ and $(q,n)$ and
using the triangle inequality, which charges the query term twice. $\square$

\section{InfoQ-X adaptation: estimator stability validation}\label{app:infoqx}

InfoQ's full criterion combines the information flow from the input to an observer point with
the information the observer carries about a per-example class target, and selects observer
points by correlating each layer's low-bit accuracy drop with its change in sliced mutual
information. Dense retrieval supplies pairwise relevance rather than a per-example class
label, so the $I(L;Y)$ branch and the accuracy-based observer selection have no canonical
one-to-one port; we evaluate the directly transferable information-flow component and refer to
the result as an \emph{InfoQ-X adaptation} rather than a reproduction. Everything outside the
sensitivity metric --- quantizer, allocation units, candidate bit widths, packed-byte budget,
calibration corpus and integer program --- is byte-for-byte the configuration used by our own
method, and the estimator is InfoQ's own SMI with a KSG estimator over random one-dimensional
projections. Projections are drawn from a shared bank across the full-precision and quantized
passes, so the two SMI estimates are positively correlated and their difference has lower
variance than independent sampling would give.

An allocator consumes the \emph{ordering} of module sensitivities, so an estimator whose
ordering does not reproduce across projection draws cannot produce a meaningful allocation
regardless of how the resulting bit assignment scores. We therefore fixed a convergence
criterion before running: two consecutive rungs of the ladder in Table~\ref{tab:infoqx} must
each show a seed-pairwise Spearman median of at least $0.9$ together with either byte-identical
selected plans or a median plan Hamming distance of at most $1$. The ladder increases
projection count first at fixed sample size, then sample count at the largest projection
setting, and stops at 6{,}000 calibration samples --- the search budget reported in the
original work. \textbf{No retrieval metric is computed during this stage}: the stability gate
is resolved first, and only an accepted estimator proceeds to a held-out NDCG comparison. This
ordering exists so that a poor baseline score cannot be explained away after the fact as
estimator noise, and equally so that a stable estimator that simply loses is reported as a
proper baseline result. \infoqxoutcome

\paragraph{Two qualifications from the implementation audit.} First, the pre-registered
protocol states that a seed redraws only the projection bank. In the code that produced
Table~\ref{tab:infoqx} the seed also redrew \emph{which} $N$ of the cached calibration
examples entered the estimate, so the rows with $N<6{,}000$ measure projection noise
\emph{and} calibration-resampling noise together, and their downward trend as $N$ grows should
not be read as a clean statement about projections. The $N=6{,}000$ row is immune --- it uses
every cached example, so the seed there changes the projection bank and nothing else --- and it
is the row the rejection rests on; the reported ceiling of \dinfoqxbestrho\ is a maximum over
all rows, so the immune row's own value cannot exceed it. The code was corrected to the
pre-registered protocol, now stores the per-seed sensitivity vectors, and the full ladder was
re-run on one development model (E5-base) as a development-only audit that does not reopen
the pre-registered decision: the maximum seed-pairwise $\rho$ over the seven rungs moved from
\diqorigmax\ to \diqfixmax, the $N=6{,}000$ row reproduced its original value exactly
(\diqorigceil\ against \diqfixceil), and no rung produced a byte-identical plan across
seeds. Second, the estimator receives the one-dimensional projections unstandardized. KSG's
joint radius is a Chebyshev distance, so a scale mismatch between the full-precision and
quantized projections can move the estimate for reasons unrelated to information. The same
audit re-ran the ladder with the projections $z$-standardised before the estimator: the
maximum $\rho$ is \diqstdmax\ and the $N=6{,}000$ row \diqstdceil, again with no reproduced
plan. Neither correction approaches the pre-registered $0.9$, so both qualifications are now
closed: what was rejected is \emph{our adaptation} of InfoQ's estimator under this protocol,
the rejection is not an artefact of either implementation choice, and we still do not claim
that the estimator family is unusable here. The more precise reading of the ladder is that the
true perturbation of information flow at three to four bits is small relative to the
estimator's finite-sample noise floor at the sampling scale the original work used.

\begin{table*}[t]\centering\footnotesize
\begin{tabular}{rrrrrr}
\toprule
Samples $N$ & Projections & Median seed $\rho$ & Median Hamming\,/12 & \textbf{Exact-plan rate} & Runtime \\
\midrule
1024 & 64 & 0.309 & 7.0 & \textbf{0\,\%} & 4\,min \\
1024 & 128 & 0.390 & 7.0 & \textbf{0\,\%} & 9\,min \\
1024 & 256 & 0.319 & 8.0 & \textbf{0\,\%} & 17\,min \\
1024 & 512 & 0.429 & 8.0 & \textbf{0\,\%} & 34\,min \\
2048 & 512 & 0.491 & 7.0 & \textbf{0\,\%} & 71\,min \\
4096 & 512 & 0.439 & 6.0 & \textbf{0\,\%} & 2.7\,h \\
6000 & 512 & 0.530 & 7.0 & \textbf{0\,\%} & 4.2\,h \\
\bottomrule
\end{tabular}

\caption{Stability of the InfoQ-X adaptation as the projection count and estimator sample
count increase, reported as the median over \dinfoqxn\ development models. Median seed $\rho$
is the seed-pairwise Spearman correlation of the $12 \times 3$ module-by-bit sensitivity
scores; median Hamming is the seed-pairwise distance between the twelve selected bit
assignments; \textbf{exact-plan rate} is the fraction of seed pairs whose selected plans are
byte-identical. \textbf{The rank correlation column is the load-bearing one.} An earlier
version of this paper emphasised the exact-plan rate instead; Appendix~\ref{app:allocsens}
shows that was a misattribution, because a well-behaved signal degraded to this level of rank
agreement also fails to reproduce an allocation. The exact-plan column is retained as a
description of what the ladder did, not as evidence about the estimator. Runtime is the median wall clock per
model per rung on one B200. The convergence criterion and the maximum sampling budget were
fixed before these runs.}
\label{tab:infoqx}
\end{table*}

\section{Pooling the query and document sides}\label{app:pooling}

Equation~\ref{eq:drift} weights the query and document sides equally. To check what that
choice costs we measured the module drift once per model with the two sides kept separate ---
sensitivity does not depend on the budget, so one measurement serves both budget points ---
and built all four poolings from that single measurement, so any difference between the rows
of Table~\ref{tab:pooling} is a difference in the pooling rule and nothing else. Across
\dpacells\ development cells the balanced, pooled and document-only variants select the same
bit plan every time and the query-only variant differs in \dpaqdiff; the balanced and pooled
held-out scores differ by \dpadelta\ points, and the module-sensitivity rankings correlate at
$\ge \dparho$. Adopting the form the margin bound implies therefore cost no measurement and
changed no development result. The same measurement on the three untouched models, made
after the freeze and shown below the rule in the table, is less tidy: at $B_{3}$ all four
variants again coincide, but at $B_{2\leftrightarrow3}$ the balanced and pooled plans differ
on \dpauntdiff\ of the three models, with the balanced form ahead by up to \dpauntdelta\
points. The freeze fixed the balanced form before these models were opened, so the reported
confirmation is the balanced form throughout and the pooled figures here are the
counterfactual.

\begin{table}[t]\centering\footnotesize\setlength{\tabcolsep}{3.5pt}
\begin{tabular}{llrrrr}
\toprule
Model & Budget & Pooled & Balanced & Query-only & Doc-only \\
\midrule
E5-base & $B_{2.5}$ & 27.29 & 27.29 & 27.29 & 27.29 \\
E5-base & $B_{3.0}$ & 33.99 & 33.99 & 33.99 & 33.99 \\
BGE-M3 & $B_{2.5}$ & 23.75 & 23.75 & 23.75 & 23.75 \\
BGE-M3 & $B_{3.0}$ & 30.15 & 30.15 & 30.15 & 30.15 \\
EmbGemma & $B_{2.5}$ & 34.89 & 34.89 & 34.89 & 34.89 \\
EmbGemma & $B_{3.0}$ & 36.99 & 36.99 & 36.93$^{\ast}$ & 36.99 \\
Qwen3-0.6B & $B_{2.5}$ & 3.38 & 3.38 & 6.55$^{\ast}$ & 3.38 \\
Qwen3-0.6B & $B_{3.0}$ & 31.31 & 31.31 & 31.31 & 31.31 \\
SKILLRET & $B_{2.5}$ & 6.15 & 6.15 & 6.15 & 6.15 \\
SKILLRET & $B_{3.0}$ & 30.62 & 30.62 & 30.62 & 30.62 \\
\midrule
mxbai-large & $B_{2.5}$ & 28.36 & 28.84$^{\ast}$ & 28.84$^{\ast}$ & 28.36 \\
mxbai-large & $B_{3.0}$ & 36.42 & 36.42 & 36.42 & 36.42 \\
MPNet-base & $B_{2.5}$ & 6.09 & 6.09 & 6.09 & 6.09 \\
MPNet-base & $B_{3.0}$ & 29.67 & 29.67 & 29.67 & 29.67 \\
GTE-base & $B_{2.5}$ & 1.66 & 2.44$^{\ast}$ & 2.44$^{\ast}$ & 1.66 \\
GTE-base & $B_{3.0}$ & 27.85 & 27.85 & 27.85 & 27.85 \\
\bottomrule
\end{tabular}

\caption{Held-out NDCG@10 of the allocation produced by each pooling of the query and document
drift, built from one shared measurement per model. Development models above the rule,
untouched models below it. Cells where a variant's plan differs from the pooled plan are
marked $^{\ast}$.}
\label{tab:pooling}
\end{table}

\section{Reproducibility of our own signal}\label{app:driftrepro}

Table~\ref{tab:driftrepro} reports the audit described in \S\ref{sec:against}: the module
sensitivity is re-measured on \ddrseeds\ independently drawn calibration subsets and the same
three statistics as the InfoQ-X ladder are computed. Every model reproduces the selected plan
exactly. As stated there, the perturbation is not the same as redrawing a projection bank and
these numbers are not a controlled estimator-variance comparison; they bound how much of our
allocation is an accident of calibration sampling, which is the question a reader of
Table~\ref{tab:baselines} would want answered.

\begin{table*}[t]\centering\footnotesize
\begin{tabular}{lrrr}
\toprule
Model & Median seed-pair $\rho$ & Median plan Hamming\,/12 & Exact-plan rate \\
\midrule
E5-base & 0.9997 & 0.0 & 100.0\% \\
BGE-M3 & 0.9997 & 0.0 & 100.0\% \\
EmbGemma & 0.9997 & 0.0 & 100.0\% \\
Qwen3-0.6B & 0.9997 & 0.0 & 100.0\% \\
SKILLRET & 0.9997 & 0.0 & 100.0\% \\
\bottomrule
\end{tabular}

\caption{Drift sensitivity under calibration resampling, at $B_{3}$. Statistics are the same
three the stability ladder reports for InfoQ-X, computed over seed pairs.}
\label{tab:driftrepro}
\end{table*}

\section{Does the gain come from the signal or from the allocator?}\label{app:nulls}

Spending a fixed byte budget unevenly is itself a choice, and it is fair to ask how much of
the improvement over uniform allocation comes from the drift signal rather than from the mere
act of allocating non-uniformly. We answer it with two nulls, both computed at $B_{3}$ on all
\dnulln\ development models and both already part of the pre-registered protocol. The
\emph{permutation} null keeps the exact multiset of module sensitivities and reassigns them to
modules at random, \dnullperm\ times: the score distribution, the allocator, the budget and the
candidate bit widths are untouched, and only the correspondence between score and module is
destroyed. The \emph{random} null draws \dnullrand\ plans uniformly from the feasible set.

Table~\ref{tab:nulls} places the drift allocation in both distributions, and the two nulls
should be read separately rather than merged into one claim. Against the random null the
result is uniform: the drift allocation is at the \dnullminrand{}th percentile on every one of
the \dnulln\ models. Against the permutation null it is strong on four and ambiguous on one.

The ambiguous model deserves a sentence in the body rather than a footnote. On EmbeddingGemma
the permutation null's \emph{median} is itself positive, so on that model reallocating bytes
unevenly beats uniform allocation even after the scores have been scrambled. The drift
allocation is still at the \dnullminperm{}th percentile there, but \textbf{some of the gain
comes from heterogeneous spending itself rather than from identifying the correct sensitive
modules}, and we do not count that model as evidence that the criterion is doing the work.
\dnullpermbeats\ of \dnulln\ models behaves this way. This is the same caution the compression-frontier analysis
raises from the other direction: where a model is far from its utility cliff, many allocations
are adequate and the choice of criterion matters less.

\begin{table*}[t]\centering\footnotesize
\begin{tabular}{lrrrr}
\toprule
Model & Drift gain (pp) & Permuted median (pp) & Permuted pct.\ & Random pct. \\
\midrule
E5-base & 1.051 & -14.776 & 100.0 & 100.0 \\
BGE-M3 & 2.478 & -25.826 & 99.5 & 100.0 \\
EmbGemma & 1.478 & 0.764 & 89.5 & 100.0 \\
Qwen3-0.6B & 0.301 & -22.372 & 100.0 & 100.0 \\
SKILLRET & 1.607 & -17.491 & 100.0 & 100.0 \\
\bottomrule
\end{tabular}

\caption{Signal against allocator structure at $B_{3}$. ``Permuted median'' is the median gain
over uniform of \dnullperm\ allocations built from randomly reassigned sensitivities; a
positive value means non-uniform spending helps on that model regardless of the criterion.
Percentile columns place the drift allocation within each null distribution.}
\label{tab:nulls}
\end{table*}

\section{How much score agreement does a reproducible allocation need?}\label{app:allocsens}

A gate that rejects a baseline has to be a gate something can pass. Our stability criterion
asks for both a rank correlation and an agreeing allocation, and the second half of that
turned out to need a control: an integer program over twelve units can be brittle, and if it
is, ``the plans never agreed'' is a statement about the allocator rather than about the
estimator that fed it.

We therefore measured the allocator's own response. Taking the measured drift scores from
\dasncells\ development cells --- a signal that is not in question --- we added noise
calibrated to reach a target rank correlation, drew two independent perturbations, ran both
through the unchanged allocator at the same budget, and counted how often the two plans
agreed. Table~\ref{tab:allocsens} is the result. Agreement is essentially guaranteed above
$\rho \approx 0.9$ and collapses below it: at \dasnthreshrho\ the two plans are identical
\dasnthreshident\ of the time, and by \dasnknee\ exact agreement has already fallen under one
in ten.

Two things follow. First, the pre-registered threshold of $0.9$ was better calibrated than we
knew when we fixed it --- it sits at the knee of this curve, which is where a sensitivity
signal stops being able to determine an allocation at all. Second, and against our own earlier
framing, the InfoQ-X exact-plan rate of zero is \emph{not} distinctive. The best rank
agreement that adaptation reached is \dinfoqxbestrho; at the control point nearest to it
(\dasnmatchrho, the closest row of Table~\ref{tab:allocsens}) the control signal itself agrees
on only \dasnmatchident\ of plan pairs, with a median Hamming distance of \dasnmatchham. The InfoQ-X
rejection rests on the rank correlation, which never came close to the threshold; the plan
statistic is a consequence of operating in a regime where no signal would reproduce an
allocation.

\begin{table*}[t]\centering\footnotesize
\begin{tabular}{rrrr}
\toprule
Score rank agreement $\rho$ & Pairs & Median plan Hamming\,/12 & Exact-plan rate \\
\midrule
0.930 & 16 & 0.0 & 75.0\% \\
0.875 & 16 & 0.0 & 56.2\% \\
0.730 & 16 & 3.0 & 6.2\% \\
0.589 & 16 & 4.0 & 6.2\% \\
0.319 & 16 & 6.0 & 0.0\% \\
0.064 & 16 & 8.0 & 0.0\% \\
0.102 & 16 & 7.0 & 0.0\% \\
\bottomrule
\end{tabular}

\caption{Allocator response to score perturbation, measured on the drift scores of
\dasncells\ development cells at $B_{3}$. Each row perturbs the real scores to a target rank
correlation, draws two independent perturbations, and compares the two resulting bit plans ---
the same quantity the stability ladder reports for InfoQ-X. Exact-plan agreement is a step
function of rank agreement, not a property of any particular estimator.}
\label{tab:allocsens}
\end{table*}

\section{Is twelve allocation units an arbitrary choice?}\label{app:granularity}

The twelve units --- four attention layer bands, four FFN layer bands, four vocabulary bands
--- were chosen to keep the configuration space enumerable, and a reader is entitled to ask
whether the results are an artifact of that choice. LieQ is natively layer-wise, interaction
effects can be hidden by aggregation, plan stability is easier to achieve over fewer units,
and the local-search topology depends on unit count, so the question touches four different
claims in the paper. We therefore re-ran the whole measurement at six, twelve and
twenty-four units on \dgrantag, holding every other setting fixed. This is a robustness
check on one development model, not a tuning sweep, and it is deliberately not run on the
untouched confirmation models.

The sensitivity ordering is essentially granularity-invariant. Projected onto a common
two-bands-per-family frame so the three settings are comparable, the pairwise rank
correlation between granularities is at least \dgranrho. Whatever the criterion is measuring,
it is not an artifact of how finely the model is cut. Retention does depend on the unit count,
but not in the direction that would embarrass the choice: six units is too coarse to beat
uniform allocation at all (\dgrangainsix\ points), twelve gains \dgrangaintwelve\ points, and
twenty-four gains \dgrangaintwentyfour\ --- less than twelve, at \dgrancost$\times$ the six-unit probe
cost, because the budget is spread over units too small for the integer program to place
usefully. Probe-to-probe transport is above $0.99$ at all three settings. Twelve is therefore
not a tuned optimum but the coarsest setting that is fine enough to help, and the criterion's
ordering would survive a different choice.

\begin{table*}[t]\centering\footnotesize
\begin{tabular}{rrrrrr}
\toprule
Units & Single-quantize evals & Probe time & Retention & Gain over uniform (pp) & Probe transport $\rho$ \\
\midrule
6 & 18 & 2.3\,min & 0.970 & -0.072 & 0.990 \\
12 & 36 & 4.4\,min & 0.983 & 0.397 & 0.994 \\
24 & 72 & 8.6\,min & 0.980 & 0.276 & 0.990 \\
\bottomrule
\end{tabular}

\caption{Allocation-unit granularity on \dgrantag\ at $B_{3}$. Probe time is the cost of
measuring module sensitivity, which grows linearly in the number of units. Retention is
held-out NDCG@10 divided by full precision. Probe transport $\rho$ is the rank correlation
of the sensitivities measured on two different probe corpora at that granularity. The
cross-granularity ordering correlations are in the text; they are computed in a common
coarse frame because the unit names differ between settings.}
\label{tab:granularity}
\end{table*}

\section{Implementation audit}\label{app:audit}

Two passes over the implementation were made after the development results were in and
before the untouched models were opened; the second pass covered the statistical primitives,
the monotone repair applied to sensitivities before allocation, and the InfoQ-X estimator. It
found three genuine bugs, all in rank statistics, and none that changes a reported
conclusion. The Spearman routine broke ties by array position rather than by average rank,
which also returns a perfect correlation for a constant vector; the Kendall routine used a
tie-corrected denominator that is wrong whenever a pair is tied in both arguments; and the
plan Hamming distance iterated over one plan's keys only. All three were corrected to agree
with the reference implementations, and every correlation in this paper whose inputs are
stored in a ledger --- \dauditn\ statistics across the predictability pools, the random-only
pools, the pooling and transport agreements and the granularity checks --- was recomputed
with the corrected routines. The largest change is \dauditmaxd\ in $\rho$, \dauditcross\ of
them cross a pre-registered threshold, and no input vector is constant. The InfoQ-X ladder's
seed-pairwise correlations cannot be recomputed because the original runs stored the
statistics rather than the per-seed vectors; the old routine inflates agreement when both
vectors are tied at the same positions and does not deflate it otherwise, so the corrected
values can only be lower. The criterion-agreement statistic $A(c)$ is defined on the raw
module drift while the allocator consumes the monotone-repaired scores; recomputing $A(c)$
on the repaired bit-3 slice gives the same value in all \dpacells\ development cells, because
the repair never touched that slice. The qualifications to the InfoQ-X result are stated in
Appendix~\ref{app:infoqx}. After the confirmation results were in, the three untouched
$B_{3}$ cells were re-evaluated with the ledger's own bit plans on a different machine
(Table~\ref{tab:reeval}); every arm agreed with the ledger to within \dreevalmaxd\ points,
and on mxbai-large the LieQ-abs pipeline --- a separate script --- had already produced the
uniform plan byte-for-byte with an NDCG identical to the uniform arm of the search pipeline
to four decimals, which is the cross-path control for the uniform evaluation. The audit
ledger, the corrected routines and their tests are in the released code.

\section{Supporting tables}\label{app:tables}

\begin{table*}[t]\centering\footnotesize
\begin{tabular}{lcccc}
\toprule
Model & Pre-specified & LieQ-native alloc. & Post-hoc best & Arms \\
\midrule
E5-base & 24.40 & 2.02 & 24.40 & 15 \\
BGE-M3 & 0.96 & 1.81 & 1.81 & 15 \\
EmbGemma & 36.13 & 9.79 & 36.86 & 15 \\
Qwen3-0.6B & 1.31 & 1.11 & 1.31 & 15 \\
SKILLRET & 1.20 & 1.82 & 1.82 & 15 \\
\bottomrule
\end{tabular}

\caption{LieQ-adapt variants at $B_{3}$. ``Pre-specified'' is the arm fixed in advance and
reported in Table~\ref{tab:baselines}. ``LieQ-native alloc.'' applies LieQ's own two-level
top-$m$ rule at a matched packed budget; over twelve coarse units a binary four-/two-bit split
cannot spend the budget as precisely as the integer program can, which is a consequence of the
granularity mismatch rather than of the criterion. ``Post-hoc best'' selects the highest
held-out score among all arms and is therefore an \emph{upper bound for the LieQ family}, not a
competitor --- our method is not given the same selection. The gap between the pre-specified
column and the post-hoc column measures how much of that family's headroom is invisible without
looking at the test metric.}
\label{tab:lieqoracle}
\end{table*}

\begin{table*}[t]\centering\footnotesize
\begin{tabular}{lcccc}
\toprule
Model & Uniform & Pre-specified & Post-hoc bound & Variants \\
\midrule
E5-base & 31.24 & 31.24 & 31.24 & 12 \\
BGE-M3 & 27.67 & 26.41 & 30.14 & 12 \\
EmbGemma & 36.13 & 30.82 & 35.31 & 12 \\
Qwen3-0.6B & 31.00 & 5.14 & 29.97 & 12 \\
SKILLRET & 29.01 & 4.96 & 28.10 & 12 \\
\bottomrule
\end{tabular}

\caption{Our own static-geometry operationalization at $B_{3}$, retained for reference now
that a prior geometry-based baseline (LieQ-adapt, Table~\ref{tab:baselines}) is available.
Layer importance is read from full-precision activation spectra with no quantization
intervention and routed through the same allocator. It is not a reproduction of any published
system. ``Pre-specified'' is a single variant fixed in advance and is the honest baseline;
``post-hoc bound'' is the best of the listed variants \emph{selected on the held-out set},
which makes it an upper bound for the family rather than a competitor --- our method is not
given the same advantage. Unlike LieQ, these proxies read the spectrum in absolute terms, with
no untrained reference, so they do not measure training-induced concentration.}
\label{tab:staticgeo}
\end{table*}

\begin{table*}[t]\centering\footnotesize
\begin{tabular}{lccccc}
\toprule
Model & All & Top 50\% & Top 25\% & Top 10\% & Within $\pm2$ of uniform \\
\midrule
E5-base & 0.892 & 0.588 & 0.328 & 0.615 & 0.698 \\
BGE-M3 & 0.973 & 0.861 & 0.664 & 0.800 & 0.762 \\
EmbGemma & 0.891 & 0.826 & 0.762 & 0.142 & 0.822 \\
Qwen3-0.6B & 0.916 & 0.581 & 0.728 & 0.022 & 0.663 \\
SKILLRET & 0.884 & 0.552 & 0.709 & 0.792 & 0.735 \\
\midrule
Macro & 0.911 & 0.682 & 0.638 & 0.474 & 0.736 \\
Median $n$ & 174 & 87 & 43 & 17 & --- \\
\bottomrule
\end{tabular}

\caption{Drift-to-NDCG correlation stratified by configuration quality, with the median sample
size per band (\S\ref{sec:predict}). A single pooled $\rho$ over a pool spanning usable and
collapsed models overstates how well drift discriminates in the region an allocator actually
operates in.}
\label{tab:stratified}
\end{table*}

\begin{table*}[t]\centering\footnotesize
\begin{tabular}{lcccc}
\toprule
Cell & Pearson $r$ & Spearman $\rho$ & Median $|R|/D$ & P90 $|R|/D$ \\
\midrule
E5-base $B_{2\leftrightarrow3}$ & 0.917 & 0.940 & 11.2\% & 40.9\% \\
E5-base $B_{3}$ & 0.966 & 0.977 & 12.4\% & 27.2\% \\
BGE-M3 $B_{2\leftrightarrow3}$ & 0.878 & 0.921 & 16.1\% & 69.1\% \\
BGE-M3 $B_{3}$ & 0.883 & 0.908 & 19.6\% & 67.7\% \\
EmbGemma $B_{2\leftrightarrow3}$ & 0.949 & 0.983 & 10.4\% & 41.3\% \\
EmbGemma $B_{3}$ & 0.945 & 0.986 & 7.5\% & 25.8\% \\
Qwen3-0.6B $B_{2\leftrightarrow3}$ & 0.873 & 0.865 & 28.2\% & 56.0\% \\
Qwen3-0.6B $B_{3}$ & 0.961 & 0.970 & 14.9\% & 37.8\% \\
SKILLRET $B_{2\leftrightarrow3}$ & 0.848 & 0.865 & 40.9\% & 70.5\% \\
SKILLRET $B_{3}$ & 0.960 & 0.976 & 14.0\% & 44.1\% \\
\midrule
Macro ($n=10$) & 0.918 & 0.939 & 17.5\% & 48.0\% \\
\bottomrule
\end{tabular}

\caption{Correlation and relative residual of the additive prediction
(\S\ref{sec:composition}). $R = D_{\mathrm{whole}} - \sum_i d_{i,b_i}$ is the composition
residual and $|R|/D_{\mathrm{whole}}$ its relative magnitude, over the 200 evaluated
configurations per cell. Read alongside Table~\ref{tab:composition}, which reports the same
disagreement on scale-free axes; the relative figures here inherit a small-denominator
inflation that those do not.}
\label{tab:additivity}
\end{table*}

\begin{table*}[t]\centering\footnotesize
\begin{tabular}{llccc}
\toprule
Budget & Model & Neighbours & Beating additive & Share \\
\midrule
$B_{2\leftrightarrow3}$ & E5-base & 100 & 2 & 2\% \\
$B_{2\leftrightarrow3}$ & BGE-M3 & 100 & 11 & 11\% \\
$B_{2\leftrightarrow3}$ & EmbGemma & 85 & 0 & 0\% \\
$B_{2\leftrightarrow3}$ & Qwen3-0.6B & 100 & 21 & 21\% \\
$B_{2\leftrightarrow3}$ & SKILLRET & 100 & 19 & 19\% \\
$B_{3}$ & E5-base & 100 & 0 & 0\% \\
$B_{3}$ & BGE-M3 & 6 & 2 & 33\% \\
$B_{3}$ & EmbGemma & 60 & 4 & 7\% \\
$B_{3}$ & Qwen3-0.6B & 100 & 11 & 11\% \\
$B_{3}$ & SKILLRET & 100 & 6 & 6\% \\
\bottomrule
\end{tabular}

\caption{How many budget-preserving neighbours of the additive plan actually beat it, by
measured held-out NDCG@10 (\S\ref{sec:search}). Where the count is zero, no selector ---
however good --- can improve on the seed within that neighbourhood.}
\label{tab:headroom}
\end{table*}

\begin{table*}[t]\centering\footnotesize
\begin{tabular}{llccccl}
\toprule
Model & Budget & Packed bpw & FP32 NDCG & Uniform ret. & Drift ret. & Drift operating regime \\
\midrule
E5-base & $B_{2\leftrightarrow3}$ & 3.117 & 34.58 & 0.047 & 0.197 & collapse \\
E5-base & $B_{3}$ & 3.612 & 34.58 & 0.904 & 0.934 & useful \\
BGE-M3 & $B_{2\leftrightarrow3}$ & 3.708 & 31.50 & 0.031 & 0.754 & useful \\
BGE-M3 & $B_{3}$ & 4.200 & 31.50 & 0.878 & 0.957 & useful \\
EmbGemma & $B_{2\leftrightarrow3}$ & 2.750 & 38.52 & 0.039 & 0.879 & useful \\
EmbGemma & $B_{3}$ & 3.250 & 38.52 & 0.938 & 0.964 & useful \\
Qwen3-0.6B & $B_{2\leftrightarrow3}$ & 3.500 & 34.97 & 0.028 & 0.097 & collapse \\
Qwen3-0.6B & $B_{3}$ & 4.000 & 34.97 & 0.887 & 0.895 & useful \\
SKILLRET & $B_{2\leftrightarrow3}$ & 3.500 & 33.74 & 0.045 & 0.182 & collapse \\
SKILLRET & $B_{3}$ & 4.000 & 33.74 & 0.860 & 0.908 & useful \\
\bottomrule
\end{tabular}

\caption{Compression frontier, with the realized packed bits per weight for each budget name
(\S\ref{sec:setup}, \S\ref{sec:against}). The realized cost of a given budget name differs
across models, which is why the names must not be read as bit widths. A cell is marked
collapse when the drift allocation retains under half of full-precision quality, and saturated
when uniform quantization already retains over 95\%, leaving little for any allocator to win.}
\label{tab:frontier}
\end{table*}

\end{document}